\documentclass[12pt]{article}    

\usepackage{latexsym} 
\usepackage{amssymb}  

\newcommand{\al}{\alpha}
\newcommand{\be}{\beta}
\newcommand{\ga}{\gamma}
\newcommand{\Ga}{\Gamma}
\newcommand{\de}{\delta}
\newcommand{\De}{\Delta}
\newcommand{\ep}{\varepsilon}
\newcommand{\eps}{\epsilon}
\newcommand{\ze}{\zeta}
\newcommand{\ka}{\kappa}
\newcommand{\la}{\lambda}
\newcommand{\La}{\Lambda}
\newcommand{\ph}{\varphi}
\newcommand{\del}{\nabla}
\newcommand{\si}{\sigma}
\newcommand{\Si}{\Sigma}
\renewcommand{\th}{\theta}   
\newcommand{\Up}{\Upsilon}
\newcommand{\om}{\omega}
\newcommand{\Om}{\Omega}

\newcommand{\imp}{~\Rightarrow}
\newcommand{\p}{\partial}
\newcommand{\<}{\langle} 
\renewcommand{\>}{\rangle} 
\newcommand{\ul}{\underline}
\newcommand{\txt}{\textstyle}
\newcommand{\dsp}{\displaystyle}
\newcommand{\h}{\hbox}
\newcommand{\ad}{\dagger}
\newcommand\eqn[1]{(\ref{#1})}      
\newcommand\Eqn[1]{Eq.~(\ref{#1})}  
\newcommand{\e}{ {\rm e} }
\newcommand{\pv}{{\bf p}}
\newcommand{\beq}{\begin{equation}}
\newcommand{\eeq}{\end{equation}}
\newcommand{\ba}{\begin{array}}
\newcommand{\bea}{\begin{eqnarray}}
\newcommand{\ea}{\end{array}}
\newcommand{\eea}{\end{eqnarray}}

\newcommand{\lab}{\label}
\renewcommand{\slash}{\!\!\!\!/\,}
\newcommand{\sslash}{\!\!\!/\,}
\newcommand{\dotprod}{\!\cdot\!}
\newcommand{\vfilll}{\vskip 0pt plus 1filll}
\newcommand{\ol}{\overline}
\newcommand{\nl}{\hfil\break}
\newcommand{\ee}[1]{\times 10^{#1}}
\newcommand\comment[1]{ \hbox{[{\it Comment suppressed here.}\/]} }
\newcommand\hide[1]{}

\renewcommand{\O}{ {\cal O} }
\newcommand{\tr}{\hbox{tr}}
\newcommand{\Tr}{\hbox{Tr}}
\newcommand{\hc}{ {\rm h.c.} }
\renewcommand{\Re}{ {\rm Re}\, }
\newcommand{\ie}{{\it i.e.}}
\newcommand{\eg}{{\it e.g.}}
\newcommand{\seq}{\!=\!}        

\newcommand{\skipover}[1]{}
\newcommand{\nn}{\nonumber \\}
\newcommand{\pl}{{\rm pl}}
\newcommand{\rt}{{\rm rt}}
\newcommand{\pg}{{\rm pg}}
\newcommand{\order}{{\cal O}}
\newcommand{\half} {{\txt {1\over 2}}}
\newcommand{\third}{{\txt {1\over 3}}}
\newcommand{\sixth}{{\txt {1\over 6}}}

\newcommand{\Dsl}{D\slash}
\newcommand{\Fmn}{F_{\mu\nu}}
\newcommand{\psibar}{{\bar\psi}}
\newcommand{\Omb}{{\bar\Om}}
\newcommand{\Csw}{C_F}
\newcommand{\csw}{c_{\rm sw}}
\newcommand{\MeV}{{\rm MeV}}
\newcommand{\GeV}{{\rm GeV}}
\newcommand{\fm}{{\rm fm}}

\def\appendix{\par                              
    \setcounter{section}{0}                     
    \setcounter{subsection}{0}
    \renewcommand{\theequation}{\Alph{section}.\arabic{equation}}
    \renewcommand{\thesection}{Appendix \Alph{section}
                \setcounter{equation}{0}  } 
}

\catcode`\@=11

\def\applabel#1{\@bsphack
  \protected@write\@auxout{}%
         {\string\newlabel{#1}{{\Alph{section}}{\thepage}}}%
  \@esphack}

\def\section{
\setcounter{equation}{0}        
\@startsection {section}{1}{\z@}{-3.5ex plus -1ex minus 
 -.2ex}{2.3ex plus .2ex}{\large\bf}}
\renewcommand{\theequation}{\arabic{section}.\arabic{equation}}

\def\subsection{\@startsection{subsection}{2}{\z@}{-3.25ex plus -1ex minus 
 -.2ex}{1.5ex plus .2ex}{\normalsize\bf}}

\def\subsubsection{\@startsection{subsubsection}{3}{\z@}{-3.25ex plus
 -1ex minus -.2ex}{1.5ex plus .2ex}{\normalsize}}

\def\@eqnnum{%
\savebox{\eqnumb}{\rm (\theequation)}%
\settowidth{\numblen}{\usebox{\eqnumb}}%
\makebox[\numblen][l]{\usebox{\eqnumb}~~~\usebox{\eqlabel}}}

\catcode`\@=12

\newsavebox{\eqlabel}

\catcode`\@=11
\newlength{\numblen}
\newsavebox{\eqnumb}
\def\@eqnnum{%
\savebox{\eqnumb}{\rm (\theequation)}%
\settowidth{\numblen}{\usebox{\eqnumb}}%
\makebox[\numblen][l]{\usebox{\eqnumb}~~~\usebox{\eqlabel}}%
}
\catcode`\@=12

\newenvironment{equationwithlabel}[1]{ %
  \savebox{\eqlabel}{#1}
   \[ \label{#1} } { \]}\savebox{\eqlabel}{~}
\newcommand{\beql}[1]{\begin{equationwithlabel}{#1}}
\newcommand{\eeql}{\end{equationwithlabel}}

\newenvironment{eqnarraywithlabel}[1]{ %
  \savebox{\eqlabel}{#1}
  \begin{eqnarray}\label{#1} }{\end{eqnarray}\savebox{\eqlabel}{~}}

\newcommand{\beal}[1]{\begin{eqnarraywithlabel}{#1}}
\newcommand{\eeal}{\end{eqnarraywithlabel}}
\begin{document}

\title{ A quark action for very coarse lattices }

\newcommand{\ns}{\normalsize}

\author{
\begin{tabular}{c@{\protect\phantom{XXXXXXX}}c}
        Mark Alford                     & Timothy R. Klassen \\[0.5ex]
        \ns School of Natural Sciences  & \ns SCRI \\
        \ns Institute for Advanced Study& \ns Florida State University \\
        \ns Princeton, NJ 08540         & \ns Tallahassee, FL 32306
\end{tabular}
\\[6ex]
        G. Peter Lepage \\[0.5ex]
        \ns Lab of Nuclear Studies \\
        \ns Cornell University \\
        \ns Ithaca, NY 14853 \\[2ex]
}

\newcommand{\preprintno}{
  \normalsize 
}

\date{\today \\ \preprintno}

\begin{titlepage}
\maketitle
\def\thepage{}          

\begin{abstract}

We investigate a tree-level  $\O(a^3)$-accurate
action, D234c, on coarse lattices. For the improvement terms we use
tadpole-improved coefficients, with the tadpole contribution measured
by the mean link in Landau gauge.

We measure the hadron spectrum for quark masses near that of the
strange quark. We find that D234c shows much better rotational
invariance than the Sheikholeslami-Wohlert action, and that mean-link
tadpole improvement leads to smaller finite-lattice-spacing errors
than plaquette tadpole improvement.
We obtain accurate ratios of lattice spacings
using a convenient ``Galilean
quarkonium'' method.

We explore the effects of possible $\O(\al_s)$ changes to the
improvement coefficients, and find that the two leading coefficients
can be independently tuned: hadron masses are most sensitive to the
clover coefficient $\Csw$, while hadron dispersion relations are most
sensitive to the third derivative coefficient $C_3$.  Preliminary
non-perturbative tuning of these coefficients yields values that are
consistent with the expected size of perturbative corrections.

\end{abstract}

\end{titlepage}

\renewcommand{\thepage}{\arabic{page}}


\section{Introduction}
\newcommand{\lag}{{\cal L}}
Lattice QCD remains the only complete implementation of
nonperturbative QCD and so is essential for low-energy QCD
phenomenology. However, simulations of lattice QCD rely upon brute
force Monte Carlo evaluations of the QCD path integral, and are
very costly. In recent years it has been demonstrated that this cost
is dramatically reduced by using coarse lattices, with lattice
spacings as large as $a\!=\!0.4$\,fm, together with more accurate
discretizations of QCD.  While highly corrected discretizations of
gluon and heavy-quark actions are now commonplace, 
less progress has been made with the much harder
problem of constructing highly improved light-quark actions.
The best light-quark actions in widespread use have finite-$a$ errors
proportional to $a^2$, which are large compared with the
$a^4,\alpha_s^2 a^2$ errors for improved gluon actions. The problem is
compounded by the fact that the effective lattice spacing for light
quarks is $2a$ rather than $a$, because the light-quark action, unlike
the others, involves first-order derivatives. 
There are a number of problems, like relativistic heavy-quark physics
and high-momentum form factors, where $\O(a^2)$ improvement is
crucial (in particular when used in conjunction with anisotropic
lattices)  for accurate results without more or less uncontrolled 
extrapolations over large mass and/or momentum regions.
In this paper we take a step towards remedying
this situation by presenting new results obtained using a
highly corrected lattice action for light-quarks.

The finite-$a$ errors can be removed, order-by-order in $a$,
from a lattice lagrangian by adding correction terms:
\begin{equation}
\lag = \lag_0 + \sum_i a^{n_i} c_i \lag_i.
\end{equation}
In principle, the coefficients $c_i$ of these correction terms can be
computed using (weak-coupling) perturbation theory, but 
in lattice QCD there has been a long-term 
reluctance to rely on perturbation
theory for any of the ingredients in QCD simulations. For most of the
past twenty years this has meant that no correction terms were
included in the action, which then has only one parameter, the bare
quark mass; the mass is tuned nonperturbatively to give correct hadron
masses. Recently a practical technique has been developed for
nonperturbatively computing the coefficient of the
$\order(a)$~correction\cite{LPCAC}. 
The $\order(a)$-accurate quark action, originally
discussed in~\cite{SW}, has led to substantial improvements over past
work, but it is still of limited value for lattice spacings larger than
0.1--0.2\,fm. 

With only a few exceptions, it is very difficult to compute
the coefficients for $\order(a^2)$ and higher corrections
nonperturbatively. Thus a perturbative
determination is the only practical alternative that permits further
improvement. Given the advantages of very coarse lattices, we feel it
is too restrictive to abandon perturbation theory completely. This is
particularly the case since we now know that perturbation theory is
generally quite reliable, provided one uses tadpole-improved lattice
operators\,\cite{LM,gplTsukuba}. In particular, perturbation theory
correctly {\em predicted\/} the relatively large renormalization of
the $\O(a)$~correction to the quark action several years before it was
confirmed in nonperturbative studies.

The coupling constant, $\alpha_s(\pi/a)$, is larger on coarser
lattices, and therefore perturbation theory is less convergent. This
makes perturbation theory less practical for calculating such things
as the overall renormalization factors relating lattice currents to
continuum currents. The correction terms in the quark action, however,
are suppressed by explicit powers of the lattice spacing. Consequently
they require less high precision, and even low-order perturbation
theory may suffice for results accurate at the few percent level.

In this paper we derive a tadpole-improved
$\order(a^3)$-accurate quark action, ``D234c''. We compare its
predictions with the those of the standard Sheikholeslami-Wohlert (SW)
$\order(a)$-accurate action, and also with the original Wilson (W)
action. To study finite-lattice-spacing errors it is not necessary to
take the chiral limit, so
we restrict our study to quark masses near the strange quark's
mass. Since finite-$a$ errors tyically grow with quark mass, our
results should improve for $u$~and $d$~quarks.

The important points in our analysis are:
\begin{itemize}

\item We use the mean link in Landau gauge rather than the traditional
plaquette prescription for calculating our tadpole improvement factor
$u_0$. Our reasons are: (1) it gives a more rotationally invariant
static potential \cite{OurPotl}; (2) it has been shown in NRQCD that
it leads to smaller scaling errors in the charmonium hyperfine
splitting \cite{Trottier}; (3) for Wilson glue, it gives a clover
coefficient that agrees more closely with the non-perturbatively
determined value \cite{gplTsukuba}. These studies suggest
that the mean-link tadpole prescription has smaller quantum
corrections than the plaquette prescription.  Of course, once higher
order perturbative corrections are included the two prescriptions will
come into agreement \cite{gplTsukuba}.

\item After tadpole improvement, the improvement coefficients are
expected to have quantum corrections of order $\al_s$, which is $\sim
0.4$ on our coarsest lattice.  In Monte-Carlo simulations, we
systematically study the effects of corrections of this size, and find
that the clover coefficient $\Csw$ is the only one whose quantum
corrections will affect hadron masses significantly, and the third
derivative coefficient $C_3$ is the only one that affects hadron
dispersion relations significantly.

\item We perform various non-perturbative tests of the coefficients of
the improvement terms. We measure the hadron dispersion relation
(``speed of light'') to check the $a^2\De^{(3)}$ term; vector meson
($\phi$) scaling as a check on the relative weight of the clover and
Wilson terms; $r$-dependence as an additional check on the relative
weight of the Wilson and clover terms and the effects of ghost
branches in the quark dispersion relation.

\item
We perform a rough non-perturbative tuning of the two leading
coefficients of the $D234$ action, and discuss the comparison with
perturbative expectations.

\item We set our overall scale from the charmonium $P-S$ splitting.
However for comparisons of scaling it is ratios of lattice spacings
that are important, and we introduce a simple method for determining
these more accurately, and with less vulnerability to systematic
errors.  It consists of measuring $P-S$ in a fictitious heavy quark
``Galilean quarkonium'' state, i.e.~using an NRQCD heavy quark action in
which relativistic corrections are not included.  Varying the quark
mass changes the size of the state, and thereby tests for the presence
of finite-$a$ errors.

\end{itemize}

We have previously studied a plaquette-tadpole-improved $\O(a^2)$-accurate
action on isotropic lattices \cite{LAT95}, and plaquette-tadpole-improved
$\O(a^3)$-accurate actions on anisotropic lattices
\cite{LAT96,Improving}, and found good dispersion relations
and scaling of mass ratios.
In this paper we find that mean-link tadpole improved D234c has the
same benefits, plus much smaller finite-$a$ errors
in  hadron masses. This is as expected,
because mean-link tadpole improvement gives a larger clover coefficient.

\section{D234c Quark action}
Following \cite{Improving},
we construct a quark action that is continuum-like 
(at tree level) through $\O(a^3)$. We start with
the continuum quark action:
\beq 
S = \psibar_c M_c \psi_c, \qquad
M_c = \Dsl + m_c 
\eeq
If we discretize this directly, our quark dispersion relation will
contain unwanted doublers at the edges of the Brillouin zone.
To avoid this, we perform a field redefinition, parameterized
by $r$, before discretizing:
\beq
\ba{rcl}
\psi_c &=& \Om\,\psi \\[1ex]
\psibar_c &=& \psibar \,\Omb \\
\ea
\eeq
where 
\beq
\Om^2 = \Omb^2 = 1 - \half r a_t( D\slash - m_c ).
\eeq
Now $S = \psibar M \psi$, where the transformed continuum quark operator is
\beq
 M = \Omb M_c \Om =  \Dsl + m_c - \half r a (\Dsl^2 - m_c^2).
\eeq
We use
$ D\slash^2 = \sum_\mu D^2_\mu - \half \si\cdot F$,
and discretize, allowing errors of order $a^4$, to obtain the
lattice D234c quark action:
\beq
\label{action}
\ba{rcl}
 M_{{\rm D234c}} &=& \dsp m_c (1 + {1\over 2} r a m_c) 
 +\sum_\mu \Bigl\{
 \ga_\mu \De^{(1)}_\mu  - {C_3\over 6} a^2 \ga_\mu\De^{(3)}_\mu \Bigr. \nn
 & + & \dsp r \Bigl. \Bigl[  - 
 {1\over 2} a  \Delta^{(2)}_\mu
 - {\Csw\over 4} a \sum_{\nu} \sigma_{\mu\nu}  \Fmn
 + {C_4 \over 24} a^3 \De^{(4)}_\mu  \Bigr]\Bigr\}~.
\ea
\eeq
$\De^{(n)}_\mu$ is the most local centered lattice discretization of the 
gauge-covariant $n$'th derivative \cite{Improving,gplSchladming};
$\De^{(3)} = \De^{(1)}\De^{(2)} = \De^{(2)}\De^{(1)}$, and
$\De^{(4)} = \De^{(2)}\De^{(2)}$.
The field strength consists of the standard clover term 
$\Fmn^{(cl)}$,
and a relative $\O(a^2)$ correction
\cite{Improving} with coefficient $C_{2F}$:
\beq
 \Fmn(x) \equiv  \Fmn^{(cl)}(x) \, - \, a^2{C_{2F}\over 6} \, 
         ( \De^{(2)}_\mu \,+\,  \De^{(2)}_\nu ) \, \Fmn^{(cl)}(x) \nn
\eeq

At tadpole-improved tree level, all links are divided by the
Landau gauge mean link $u_0$, and $\Csw=C_3=C_4=C_{2F}=1$.
We will explore the effects of deviations from these values.

The terms proportional to $r$ remove the doublers from the quark
dispersion relations, so that for generic values of $r \sim 1$ this is
a doubler-free tree-level $\O(a^3)$-accurate quark action.  The
derivation can be straightforwardly generalized to anisotropic
lattices \cite{Improving,gplSchladming}.

\begin{figure}[Htb]
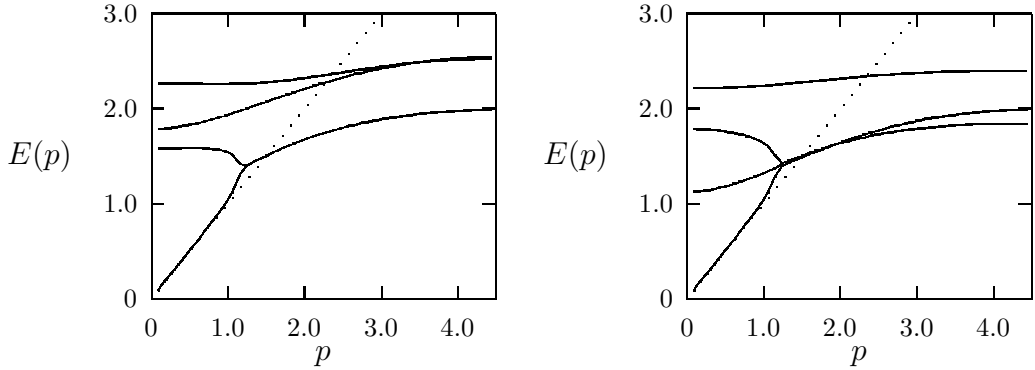

\begin{center}
\setlength{\unitlength}{0.240900pt}
\ifx\plotpoint\undefined\newsavebox{\plotpoint}\fi
\sbox{\plotpoint}{\rule[-0.200pt]{0.400pt}{0.400pt}}%


\end{center}
\caption{ 
Massless dispersion relation ${\rm Re}[ a E(a|{\bf p}|) ]$ 
for D234c($r\!=\!1$) and D234c($r\!=\!2/3$) quarks on an isotropic lattice.
Note the continuum-like behavior at low momentum, and the presence
of three ghost branches, one of which drops dramatically
as $r$ changes from 1 to $2/3$.
}
\label{fig:disprel1}
\end{figure}

For $r=1$ there are three fairly high ghost branches in the free
quark dispersion relation (Figure \ref{fig:disprel1}).
To investigate the effect of redundant terms
we will also study $r=2/3$,
for which one of the ghost branches moves down so that $E(0)\approx 1.0$.

Note that the 
two leading terms both violate symmetries that will be restored in the
continuum limit, and hence can be non-perturbatively tuned.
The clover term violates chiral symmetry, and so can be
tuned by imposing PCAC \cite{LPCAC}.
The $\De^{(3)}$ term is the {\em only} rotational symmetry-violating
$\O(a^2)$ term, and so its coefficient $C_3$
can be nonperturbatively tuned by imposing rotational invariance.

\section{Gluon action and lattice spacing determination}
\label{sec:lattspac}

We use a tree-level tadpole-improved plaquette and 
$2\!\times\!1$ rectangle glue action \cite{WW,LW,Alf1},
\beq
\ba{rcl}
 S &=&\dsp - \beta \sum_{x,\mu}
\left\{ 
\frac{5}{3}  \frac{P_{\mu\nu}(x)}{u_0^4}
- \frac{1}{12} \frac{R_{\mu\nu}(x)}{u_0^6}
- \frac{1}{12} \frac{R_{\nu\mu}(x)}{u_0^6}
\right\}, \\[3ex]
\quad P_{\mu\nu} 
 &=&\dsp \third{\rm Re}\Tr W(\mu+\nu-\mu-\nu) \quad \hbox{(plaquette)}, \\[1ex]
R_{\mu\nu} 
  &=&\dsp \third{\rm Re}\Tr W(2\mu+\nu-2\mu-\nu) \quad \hbox{(rectangle)},
\ea
\eeq
where a Wilson loop $W(\rho-\si\cdots)$ goes one link in the $\rho$
direction, one link in the negative $\si$ direction, etc.

This definition of $\beta$ is different from that of
Ref.~\cite{Alf1}, where a factor of~$5/3u_0^4$ was
absorbed into~$\beta$. We prefer the notation here because
$\beta\!=\!6/g^2$ as in the original Wilson action. Furthermore the coupling
\beq\label{alpha}
\alpha_s \equiv {3\over 2\pi\beta}
\eeq
is now tadpole-improved and
therefore roughly equal to continuum couplings like
$\alpha_V(\pi/a_s)$.

The tadpole improvement factor $u_0$ is the mean of the link operator
in Landau gauge. 
At both our lattice spacings we found that an $8^4$ lattice was large enough
for finite volume effects in $u_0$ to be of order $0.1\%$.
To fix to Landau gauge we maximize \cite{gplTsukuba}
\begin{equation}
\sum_{x,\,\mu} \frac{1}{u_0 a^2}\,{\rm ReTr}\left\{
U_\mu(x) - \frac{1}{16u_0}\,U_\mu(x)\,U_\mu(x+\hat\mu)\right\}.
\end{equation}

We generated gluon configurations at two lattice spacings, $0.40~\fm$
and $0.25~\fm$ (see Table \ref{tab:glue}). 
The lattice spacing was determined in two ways.

Firstly, we performed NRQCD simulations of charmonium, using the
experimental value of $458~\MeV$ for the spin-averaged $P-S$
splitting; the results are given in Table \ref{tab:glue} column 6
(details in Table \ref{tab:charmglue}).  Note that the errors quoted are
statistical, and do not reflect systematic uncertainties such as
quenching, finite-$a$ errors, or higher-order relativistic effects
neglected in our NRQCD simulation. This lattice spacing determination
is therefore not suitable for precise comparisons with data from
other groups.

Secondly, a more accurate determination of {\em ratios} of lattice
spacings is possible, since there is no need to simulate a known
physical state. At each lattice spacing we can measure the mass of a
fictitious state, whose properties are chosen for convenience.  We
chose ``Galilean quarkonium,'' a bound state of a quark and antiquark in
a non-relativistic world. We simulated this state using NRQCD with no
relativistic corrections.  By making the Galilean quarkonium lighter
(and hence bigger) than charmonium we
reduce the finite-$a$ errors.  In fact, we studied a range of quark
masses down to about half the charm quark mass
and found that lattice spacing ratios were all consistent with
each other, within errors.
For details see \ref{app:lattspacing} and Table \ref{tab:galileo}.  
Since our main
goal is to compare our mean-link-tadpole-improved results with 
SCRI's plaquette-tadpole-improved
results, we used Galilean quarkonium to calculate the ratios of 
our lattice spacings to the SCRI $\be=7.4$ lattice spacing. The
results are in Table \ref{tab:galileo}.

For convenience we want to give our results an absolute energy scale, 
so we take the SCRI $\be=7.4$ lattice spacing to be $a^{-1}=812~\MeV$,
which corresponds to $\sqrt{\si}= 468~\MeV$ for their string tension.
This gives the final column of Table \ref{tab:glue},
which is consistent with our charmonium measurements.
Note that the error bars reflect the uncertainty in the
ratio to SCRI's lattice spacings, which is the relevant quantity
for scaling comparisons. It does not reflect the overall uncertainty
in the scale, which was introduced purely for convenience.

\def\st{\rule[-1.5ex]{0em}{4ex}} 
\begin{table}[ht]
\begin{tabular}{lllllll}
\hline
\st $\beta$ & $\al_s$ & $u_0$ & $u_p$ & approx $a$ & 
charmonium $a^{-1}$ & $a^{-1}$ rel.~to SCRI \\
\hline  
\st 1.157\phantom{x}
          & 0.413 & 0.738 & 0.8196 & 0.40~\fm & 495(4)~\MeV & 497(3)~\MeV \\
\st 1.719 & 0.278 & 0.797 & 0.8576 & 0.25~\fm & 790(10)~\MeV & 785(6)~\MeV \\
\multicolumn{2}{l}{7.4 (SCRI)} \st
                  &       &        & 0.24~\fm & 840(20)~\MeV & 812(def)~\MeV \\
\hline
\end{tabular}
\caption{
Parameters of glue for our two lattice spacings, and one of SCRI's.
Mean Landau-gauge link ($u_0$) was used in our tadpole improvement (TI);
plaquette TI factor $u_p = ({\rm Plaq})^{1/4}$ is given for
comparison.  The $a^{-1}$ relative to SCRI is obtained by the 
Galilean quarkonium method.
}
\label{tab:glue}
\end{table}


For hadron spectrum measurements we used lattices of the same physical
size ($2~\fm$) at both lattice spacings. We also performed a set of
measurements investigating the volume dependence of hadron masses (see
appendix, Table \ref{tab:vol}). We see that the $1.6~\fm$ lattice agrees with
the $1.75~\fm$ and $2~\fm$ lattices within statistical errors.

\def\pv{{\bf p}}
\def\csw{c_{\rm sw}}
\section{Results}

We subjected the D234c action to a series of tests to determine its
viability at large lattice spacings. We examined the scaling of the
vector meson mass and of baryon masses, and we measured hadronic
dispersion relations. We also measured the sensitivity of these
physical quantities to changes in the coefficients in the action.

\subsection{Hadron masses}

\begin{figure}[htb]
\begin{center}
\setlength{\unitlength}{0.240900pt}
\ifx\plotpoint\undefined\newsavebox{\plotpoint}\fi
\sbox{\plotpoint}{\rule[-0.200pt]{0.400pt}{0.400pt}}%
\begin{picture}(1200,1080)(0,0)
\font\gnuplot=cmr10 at 10pt
\gnuplot
\sbox{\plotpoint}{\rule[-0.200pt]{0.400pt}{0.400pt}}%
\put(176.0,68.0){\rule[-0.200pt]{0.400pt}{238.250pt}}
\put(176.0,86.0){\rule[-0.200pt]{4.818pt}{0.400pt}}
\put(154,86){\makebox(0,0)[r]{0.7}}
\put(1116.0,86.0){\rule[-0.200pt]{4.818pt}{0.400pt}}
\put(176.0,262.0){\rule[-0.200pt]{4.818pt}{0.400pt}}
\put(154,262){\makebox(0,0)[r]{0.8}}
\put(1116.0,262.0){\rule[-0.200pt]{4.818pt}{0.400pt}}
\put(176.0,439.0){\rule[-0.200pt]{4.818pt}{0.400pt}}
\put(154,439){\makebox(0,0)[r]{0.9}}
\put(1116.0,439.0){\rule[-0.200pt]{4.818pt}{0.400pt}}
\put(176.0,615.0){\rule[-0.200pt]{4.818pt}{0.400pt}}
\put(154,615){\makebox(0,0)[r]{1}}
\put(1116.0,615.0){\rule[-0.200pt]{4.818pt}{0.400pt}}
\put(176.0,792.0){\rule[-0.200pt]{4.818pt}{0.400pt}}
\put(154,792){\makebox(0,0)[r]{1.1}}
\put(1116.0,792.0){\rule[-0.200pt]{4.818pt}{0.400pt}}
\put(176.0,969.0){\rule[-0.200pt]{4.818pt}{0.400pt}}
\put(154,969){\makebox(0,0)[r]{1.2}}
\put(1116.0,969.0){\rule[-0.200pt]{4.818pt}{0.400pt}}
\put(176.0,68.0){\rule[-0.200pt]{0.400pt}{4.818pt}}
\put(176,23){\makebox(0,0){0}}
\put(176.0,1037.0){\rule[-0.200pt]{0.400pt}{4.818pt}}
\put(416.0,68.0){\rule[-0.200pt]{0.400pt}{4.818pt}}
\put(416,23){\makebox(0,0){0.05}}
\put(416.0,1037.0){\rule[-0.200pt]{0.400pt}{4.818pt}}
\put(656.0,68.0){\rule[-0.200pt]{0.400pt}{4.818pt}}
\put(656,23){\makebox(0,0){0.1}}
\put(656.0,1037.0){\rule[-0.200pt]{0.400pt}{4.818pt}}
\put(896.0,68.0){\rule[-0.200pt]{0.400pt}{4.818pt}}
\put(896,23){\makebox(0,0){0.15}}
\put(896.0,1037.0){\rule[-0.200pt]{0.400pt}{4.818pt}}
\put(1136.0,68.0){\rule[-0.200pt]{0.400pt}{4.818pt}}
\put(1136,23){\makebox(0,0){0.2}}
\put(1136.0,1037.0){\rule[-0.200pt]{0.400pt}{4.818pt}}
\put(176.0,68.0){\rule[-0.200pt]{231.264pt}{0.400pt}}
\put(1136.0,68.0){\rule[-0.200pt]{0.400pt}{238.250pt}}
\put(176.0,1057.0){\rule[-0.200pt]{231.264pt}{0.400pt}}
\put(-63,704){\makebox(0,0)[l]{\shortstack{$\phi$ mass\\ (GeV)}}}
\put(656,-37){\makebox(0,0)[l]{$a^2$ (fm${}^2$)}}
\put(176.0,68.0){\rule[-0.200pt]{0.400pt}{238.250pt}}
\put(1006,992){\makebox(0,0)[r]{D234c, mean-link TI}}
\put(1050,992){\circle*{24}}
\put(944,561){\circle*{24}}
\put(476,663){\circle*{24}}
\put(1028.0,992.0){\rule[-0.200pt]{15.899pt}{0.400pt}}
\put(1028.0,982.0){\rule[-0.200pt]{0.400pt}{4.818pt}}
\put(1094.0,982.0){\rule[-0.200pt]{0.400pt}{4.818pt}}
\put(944.0,545.0){\rule[-0.200pt]{0.400pt}{7.709pt}}
\put(934.0,545.0){\rule[-0.200pt]{4.818pt}{0.400pt}}
\put(934.0,577.0){\rule[-0.200pt]{4.818pt}{0.400pt}}
\put(476.0,646.0){\rule[-0.200pt]{0.400pt}{8.431pt}}
\put(466.0,646.0){\rule[-0.200pt]{4.818pt}{0.400pt}}
\put(466.0,681.0){\rule[-0.200pt]{4.818pt}{0.400pt}}
\put(1006,947){\makebox(0,0)[r]{SW,    mean-link TI}}
\put(1050,947){\circle{24}}
\put(944,497){\circle{24}}
\put(476,688){\circle{24}}
\put(1028.0,947.0){\rule[-0.200pt]{15.899pt}{0.400pt}}
\put(1028.0,937.0){\rule[-0.200pt]{0.400pt}{4.818pt}}
\put(1094.0,937.0){\rule[-0.200pt]{0.400pt}{4.818pt}}
\put(944.0,479.0){\rule[-0.200pt]{0.400pt}{8.672pt}}
\put(934.0,479.0){\rule[-0.200pt]{4.818pt}{0.400pt}}
\put(934.0,515.0){\rule[-0.200pt]{4.818pt}{0.400pt}}
\put(476.0,661.0){\rule[-0.200pt]{0.400pt}{12.768pt}}
\put(466.0,661.0){\rule[-0.200pt]{4.818pt}{0.400pt}}
\put(466.0,714.0){\rule[-0.200pt]{4.818pt}{0.400pt}}
\put(1006,902){\makebox(0,0)[r]{SW,  SCRI, plaq TI}}
\put(1050,902){\makebox(0,0){$\times$}}
\put(282,619){\makebox(0,0){$\times$}}
\put(314,605){\makebox(0,0){$\times$}}
\put(363,563){\makebox(0,0){$\times$}}
\put(459,496){\makebox(0,0){$\times$}}
\put(661,375){\makebox(0,0){$\times$}}
\put(927,209){\makebox(0,0){$\times$}}
\put(1028.0,902.0){\rule[-0.200pt]{15.899pt}{0.400pt}}
\put(1028.0,892.0){\rule[-0.200pt]{0.400pt}{4.818pt}}
\put(1094.0,892.0){\rule[-0.200pt]{0.400pt}{4.818pt}}
\put(282.0,598.0){\rule[-0.200pt]{0.400pt}{10.118pt}}
\put(272.0,598.0){\rule[-0.200pt]{4.818pt}{0.400pt}}
\put(272.0,640.0){\rule[-0.200pt]{4.818pt}{0.400pt}}
\put(314.0,582.0){\rule[-0.200pt]{0.400pt}{11.081pt}}
\put(304.0,582.0){\rule[-0.200pt]{4.818pt}{0.400pt}}
\put(304.0,628.0){\rule[-0.200pt]{4.818pt}{0.400pt}}
\put(363.0,550.0){\rule[-0.200pt]{0.400pt}{6.022pt}}
\put(353.0,550.0){\rule[-0.200pt]{4.818pt}{0.400pt}}
\put(353.0,575.0){\rule[-0.200pt]{4.818pt}{0.400pt}}
\put(459.0,487.0){\rule[-0.200pt]{0.400pt}{4.336pt}}
\put(449.0,487.0){\rule[-0.200pt]{4.818pt}{0.400pt}}
\put(449.0,505.0){\rule[-0.200pt]{4.818pt}{0.400pt}}
\put(661.0,365.0){\rule[-0.200pt]{0.400pt}{5.059pt}}
\put(651.0,365.0){\rule[-0.200pt]{4.818pt}{0.400pt}}
\put(651.0,386.0){\rule[-0.200pt]{4.818pt}{0.400pt}}
\put(927.0,203.0){\rule[-0.200pt]{0.400pt}{2.891pt}}
\put(917.0,203.0){\rule[-0.200pt]{4.818pt}{0.400pt}}
\put(917.0,215.0){\rule[-0.200pt]{4.818pt}{0.400pt}}
\put(1006,857){\makebox(0,0)[r]{Wil, SCRI, plaq TI}}
\put(1050,857){\makebox(0,0){$+$}}
\put(282,271){\makebox(0,0){$+$}}
\put(314,232){\makebox(0,0){$+$}}
\put(363,150){\makebox(0,0){$+$}}
\put(1028.0,857.0){\rule[-0.200pt]{15.899pt}{0.400pt}}
\put(1028.0,847.0){\rule[-0.200pt]{0.400pt}{4.818pt}}
\put(1094.0,847.0){\rule[-0.200pt]{0.400pt}{4.818pt}}
\put(282.0,252.0){\rule[-0.200pt]{0.400pt}{9.395pt}}
\put(272.0,252.0){\rule[-0.200pt]{4.818pt}{0.400pt}}
\put(272.0,291.0){\rule[-0.200pt]{4.818pt}{0.400pt}}
\put(314.0,216.0){\rule[-0.200pt]{0.400pt}{7.709pt}}
\put(304.0,216.0){\rule[-0.200pt]{4.818pt}{0.400pt}}
\put(304.0,248.0){\rule[-0.200pt]{4.818pt}{0.400pt}}
\put(363.0,140.0){\rule[-0.200pt]{0.400pt}{4.818pt}}
\put(353.0,140.0){\rule[-0.200pt]{4.818pt}{0.400pt}}
\put(353.0,160.0){\rule[-0.200pt]{4.818pt}{0.400pt}}
\sbox{\plotpoint}{\rule[-0.500pt]{1.000pt}{1.000pt}}%
\put(176,678){\usebox{\plotpoint}}
\put(176.00,678.00){\usebox{\plotpoint}}
\put(193.57,666.96){\usebox{\plotpoint}}
\multiput(195,666)(17.798,-10.679){0}{\usebox{\plotpoint}}
\put(211.32,656.21){\usebox{\plotpoint}}
\multiput(215,654)(17.270,-11.513){0}{\usebox{\plotpoint}}
\put(228.84,645.09){\usebox{\plotpoint}}
\multiput(234,642)(17.798,-10.679){0}{\usebox{\plotpoint}}
\put(246.64,634.42){\usebox{\plotpoint}}
\multiput(254,630)(17.270,-11.513){0}{\usebox{\plotpoint}}
\put(264.16,623.30){\usebox{\plotpoint}}
\put(281.56,612.01){\usebox{\plotpoint}}
\multiput(283,611)(17.270,-11.513){0}{\usebox{\plotpoint}}
\put(299.02,600.79){\usebox{\plotpoint}}
\multiput(302,599)(17.798,-10.679){0}{\usebox{\plotpoint}}
\put(316.67,589.89){\usebox{\plotpoint}}
\multiput(321,587)(17.798,-10.679){0}{\usebox{\plotpoint}}
\put(334.34,579.00){\usebox{\plotpoint}}
\multiput(341,575)(17.798,-10.679){0}{\usebox{\plotpoint}}
\put(352.10,568.27){\usebox{\plotpoint}}
\put(369.66,557.21){\usebox{\plotpoint}}
\multiput(370,557)(17.798,-10.679){0}{\usebox{\plotpoint}}
\put(387.23,546.18){\usebox{\plotpoint}}
\multiput(389,545)(17.798,-10.679){0}{\usebox{\plotpoint}}
\put(404.98,535.41){\usebox{\plotpoint}}
\multiput(409,533)(17.270,-11.513){0}{\usebox{\plotpoint}}
\put(422.30,523.99){\usebox{\plotpoint}}
\multiput(428,520)(17.798,-10.679){0}{\usebox{\plotpoint}}
\put(439.83,512.90){\usebox{\plotpoint}}
\multiput(448,508)(17.270,-11.513){0}{\usebox{\plotpoint}}
\put(457.35,501.79){\usebox{\plotpoint}}
\put(475.15,491.11){\usebox{\plotpoint}}
\multiput(477,490)(17.270,-11.513){0}{\usebox{\plotpoint}}
\put(492.67,480.00){\usebox{\plotpoint}}
\multiput(496,478)(17.798,-10.679){0}{\usebox{\plotpoint}}
\put(510.34,469.11){\usebox{\plotpoint}}
\multiput(515,466)(17.798,-10.679){0}{\usebox{\plotpoint}}
\put(527.99,458.20){\usebox{\plotpoint}}
\multiput(535,454)(17.270,-11.513){0}{\usebox{\plotpoint}}
\put(545.52,447.09){\usebox{\plotpoint}}
\put(562.90,435.77){\usebox{\plotpoint}}
\multiput(564,435)(17.798,-10.679){0}{\usebox{\plotpoint}}
\put(580.45,424.70){\usebox{\plotpoint}}
\multiput(583,423)(17.798,-10.679){0}{\usebox{\plotpoint}}
\put(598.17,413.90){\usebox{\plotpoint}}
\multiput(603,411)(17.270,-11.513){0}{\usebox{\plotpoint}}
\put(615.69,402.79){\usebox{\plotpoint}}
\multiput(622,399)(17.798,-10.679){0}{\usebox{\plotpoint}}
\put(633.44,392.04){\usebox{\plotpoint}}
\multiput(641,387)(17.798,-10.679){0}{\usebox{\plotpoint}}
\put(651.01,380.99){\usebox{\plotpoint}}
\put(668.81,370.32){\usebox{\plotpoint}}
\multiput(671,369)(17.270,-11.513){0}{\usebox{\plotpoint}}
\put(686.33,359.20){\usebox{\plotpoint}}
\multiput(690,357)(17.798,-10.679){0}{\usebox{\plotpoint}}
\put(703.80,348.05){\usebox{\plotpoint}}
\multiput(709,344)(17.798,-10.679){0}{\usebox{\plotpoint}}
\put(721.15,336.71){\usebox{\plotpoint}}
\multiput(729,332)(17.270,-11.513){0}{\usebox{\plotpoint}}
\put(738.67,325.60){\usebox{\plotpoint}}
\put(756.47,314.92){\usebox{\plotpoint}}
\multiput(758,314)(17.798,-10.679){0}{\usebox{\plotpoint}}
\put(774.08,303.95){\usebox{\plotpoint}}
\multiput(777,302)(17.798,-10.679){0}{\usebox{\plotpoint}}
\put(791.79,293.13){\usebox{\plotpoint}}
\multiput(797,290)(17.270,-11.513){0}{\usebox{\plotpoint}}
\put(809.31,282.01){\usebox{\plotpoint}}
\multiput(816,278)(17.798,-10.679){0}{\usebox{\plotpoint}}
\put(827.07,271.28){\usebox{\plotpoint}}
\put(844.63,260.22){\usebox{\plotpoint}}
\multiput(845,260)(17.004,-11.902){0}{\usebox{\plotpoint}}
\put(861.75,248.50){\usebox{\plotpoint}}
\multiput(864,247)(17.798,-10.679){0}{\usebox{\plotpoint}}
\put(879.48,237.71){\usebox{\plotpoint}}
\multiput(884,235)(17.798,-10.679){0}{\usebox{\plotpoint}}
\put(897.18,226.88){\usebox{\plotpoint}}
\multiput(903,223)(17.798,-10.679){0}{\usebox{\plotpoint}}
\put(914.80,215.92){\usebox{\plotpoint}}
\multiput(923,211)(17.270,-11.513){0}{\usebox{\plotpoint}}
\put(932.33,204.80){\usebox{\plotpoint}}
\put(950.12,194.13){\usebox{\plotpoint}}
\multiput(952,193)(17.270,-11.513){0}{\usebox{\plotpoint}}
\put(967.65,183.01){\usebox{\plotpoint}}
\multiput(971,181)(17.798,-10.679){0}{\usebox{\plotpoint}}
\put(985.44,172.33){\usebox{\plotpoint}}
\multiput(991,169)(16.383,-12.743){0}{\usebox{\plotpoint}}
\put(1002.46,160.52){\usebox{\plotpoint}}
\multiput(1010,156)(17.798,-10.679){0}{\usebox{\plotpoint}}
\put(1020.25,149.83){\usebox{\plotpoint}}
\put(1037.78,138.73){\usebox{\plotpoint}}
\multiput(1039,138)(17.798,-10.679){0}{\usebox{\plotpoint}}
\put(1055.39,127.74){\usebox{\plotpoint}}
\multiput(1058,126)(17.798,-10.679){0}{\usebox{\plotpoint}}
\put(1073.10,116.94){\usebox{\plotpoint}}
\multiput(1078,114)(17.798,-10.679){0}{\usebox{\plotpoint}}
\put(1090.82,106.12){\usebox{\plotpoint}}
\multiput(1097,102)(17.798,-10.679){0}{\usebox{\plotpoint}}
\put(1108.42,95.15){\usebox{\plotpoint}}
\put(1125.95,84.03){\usebox{\plotpoint}}
\multiput(1126,84)(17.798,-10.679){0}{\usebox{\plotpoint}}
\put(1136,78){\usebox{\plotpoint}}
\put(176,675){\usebox{\plotpoint}}
\multiput(176,675)(1.604,-20.693){7}{\usebox{\plotpoint}}
\multiput(186,546)(3.540,-20.451){2}{\usebox{\plotpoint}}
\multiput(195,494)(5.155,-20.105){2}{\usebox{\plotpoint}}
\multiput(205,455)(6.191,-19.811){2}{\usebox{\plotpoint}}
\put(219.25,409.79){\usebox{\plotpoint}}
\multiput(224,395)(7.451,-19.372){2}{\usebox{\plotpoint}}
\put(241.81,351.82){\usebox{\plotpoint}}
\put(250.65,333.04){\usebox{\plotpoint}}
\put(259.32,314.18){\usebox{\plotpoint}}
\put(268.72,295.70){\usebox{\plotpoint}}
\put(279.06,277.70){\usebox{\plotpoint}}
\put(289.07,259.53){\usebox{\plotpoint}}
\put(300.04,241.94){\usebox{\plotpoint}}
\put(311.56,224.67){\usebox{\plotpoint}}
\multiput(312,224)(11.224,-17.459){0}{\usebox{\plotpoint}}
\put(322.92,207.31){\usebox{\plotpoint}}
\put(335.18,190.56){\usebox{\plotpoint}}
\put(348.18,174.38){\usebox{\plotpoint}}
\multiput(351,171)(12.453,-16.604){0}{\usebox{\plotpoint}}
\put(360.86,157.96){\usebox{\plotpoint}}
\put(374.15,142.02){\usebox{\plotpoint}}
\put(387.36,126.01){\usebox{\plotpoint}}
\multiput(389,124)(14.676,-14.676){0}{\usebox{\plotpoint}}
\put(401.71,111.02){\usebox{\plotpoint}}
\put(415.63,95.63){\usebox{\plotpoint}}
\multiput(418,93)(14.676,-14.676){0}{\usebox{\plotpoint}}
\put(430.28,80.95){\usebox{\plotpoint}}
\multiput(438,74)(14.676,-14.676){0}{\usebox{\plotpoint}}
\put(444,68){\usebox{\plotpoint}}
\end{picture}

\end{center}
\caption{ 
Mass of $\phi$ meson as a function of lattice spacing, $P/V=0.70$.
Mean link TI ($\bullet$ D234, $\circ$ SW) clearly scales better than 
plaquette TI ($\times$ SW, + Wilson, data kindly supplied by
SCRI)
}
\label{fig:phi70}
\end{figure}
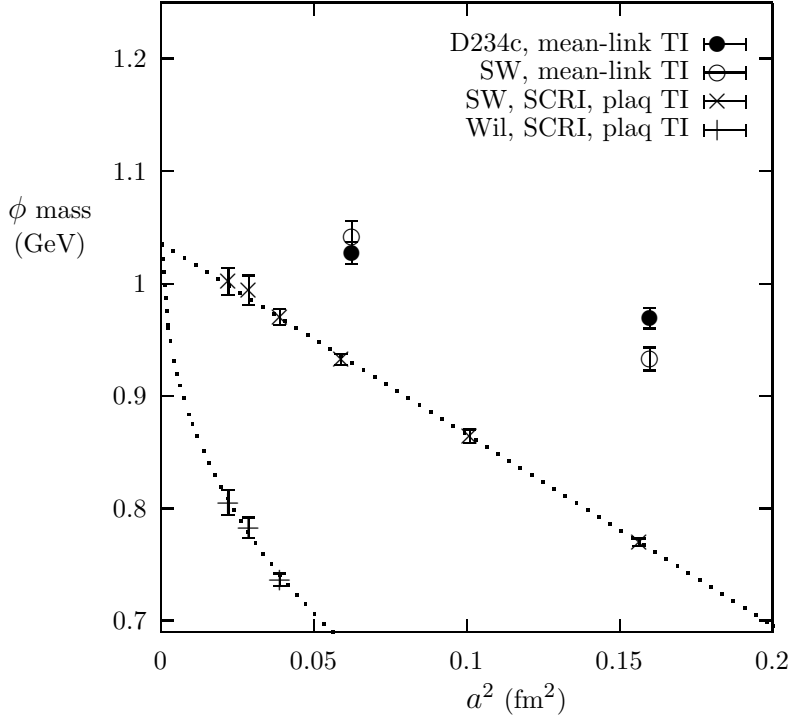

In Figure~\ref{fig:phi70} we show how the vector meson mass varies
with lattice spacing when the ratio of the pseudoscalar to vector
meson masses is $P/V=0.7$.  (Full data is in 
\ref{app:tables}, along with data for $P/V=0.76$, corresponding to a
slightly larger quark mass; see tables \ref{tab:hadrons70},
\ref{tab:hadrons76}).  
We present data obtained using the
tree-level D234c, with Landau-link tadpole improvement, at lattice
spacings~$a$ of 0.25\,fm and~0.4\,fm. These values are compared with
results from SCRI obtained using the Wilson and SW
actions\,\cite{SCRI}, which can be extrapolated to give~$a\seq0$
results, as indicated.

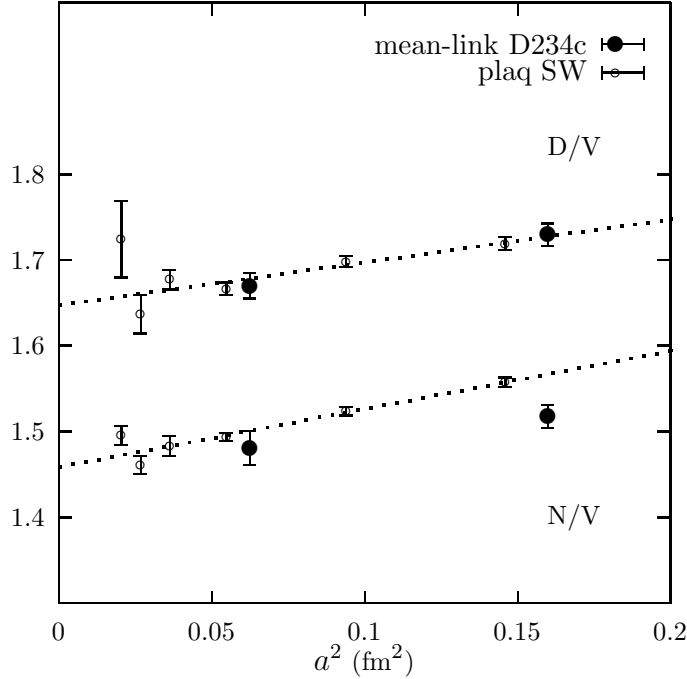
\begin{figure}[htb]
\begin{center}
\setlength{\unitlength}{0.240900pt}
\ifx\plotpoint\undefined\newsavebox{\plotpoint}\fi
\sbox{\plotpoint}{\rule[-0.200pt]{0.400pt}{0.400pt}}%
\begin{picture}(1200,1080)(0,0)
\font\gnuplot=cmr10 at 10pt
\gnuplot
\sbox{\plotpoint}{\rule[-0.200pt]{0.400pt}{0.400pt}}%
\put(176.0,113.0){\rule[-0.200pt]{0.400pt}{227.410pt}}
\put(176.0,248.0){\rule[-0.200pt]{4.818pt}{0.400pt}}
\put(154,248){\makebox(0,0)[r]{1.4}}
\put(1116.0,248.0){\rule[-0.200pt]{4.818pt}{0.400pt}}
\put(176.0,383.0){\rule[-0.200pt]{4.818pt}{0.400pt}}
\put(154,383){\makebox(0,0)[r]{1.5}}
\put(1116.0,383.0){\rule[-0.200pt]{4.818pt}{0.400pt}}
\put(176.0,518.0){\rule[-0.200pt]{4.818pt}{0.400pt}}
\put(154,518){\makebox(0,0)[r]{1.6}}
\put(1116.0,518.0){\rule[-0.200pt]{4.818pt}{0.400pt}}
\put(176.0,652.0){\rule[-0.200pt]{4.818pt}{0.400pt}}
\put(154,652){\makebox(0,0)[r]{1.7}}
\put(1116.0,652.0){\rule[-0.200pt]{4.818pt}{0.400pt}}
\put(176.0,787.0){\rule[-0.200pt]{4.818pt}{0.400pt}}
\put(154,787){\makebox(0,0)[r]{1.8}}
\put(1116.0,787.0){\rule[-0.200pt]{4.818pt}{0.400pt}}
\put(176.0,113.0){\rule[-0.200pt]{0.400pt}{4.818pt}}
\put(176,68){\makebox(0,0){0}}
\put(176.0,1037.0){\rule[-0.200pt]{0.400pt}{4.818pt}}
\put(416.0,113.0){\rule[-0.200pt]{0.400pt}{4.818pt}}
\put(416,68){\makebox(0,0){0.05}}
\put(416.0,1037.0){\rule[-0.200pt]{0.400pt}{4.818pt}}
\put(656.0,113.0){\rule[-0.200pt]{0.400pt}{4.818pt}}
\put(656,68){\makebox(0,0){0.1}}
\put(656.0,1037.0){\rule[-0.200pt]{0.400pt}{4.818pt}}
\put(896.0,113.0){\rule[-0.200pt]{0.400pt}{4.818pt}}
\put(896,68){\makebox(0,0){0.15}}
\put(896.0,1037.0){\rule[-0.200pt]{0.400pt}{4.818pt}}
\put(1136.0,113.0){\rule[-0.200pt]{0.400pt}{4.818pt}}
\put(1136,68){\makebox(0,0){0.2}}
\put(1136.0,1037.0){\rule[-0.200pt]{0.400pt}{4.818pt}}
\put(176.0,113.0){\rule[-0.200pt]{231.264pt}{0.400pt}}
\put(1136.0,113.0){\rule[-0.200pt]{0.400pt}{227.410pt}}
\put(176.0,1057.0){\rule[-0.200pt]{231.264pt}{0.400pt}}
\put(656,23){\makebox(0,0){$a^2$ (fm${}^2$)}}
\put(944,248){\makebox(0,0)[l]{N/V}}
\put(944,828){\makebox(0,0)[l]{D/V}}
\put(176.0,113.0){\rule[-0.200pt]{0.400pt}{227.410pt}}
\put(1006,992){\makebox(0,0)[r]{{\small mean-link D234c}}}
\put(1050,992){\circle*{24}}
\put(944,407){\circle*{24}}
\put(476,357){\circle*{24}}
\put(1028.0,992.0){\rule[-0.200pt]{15.899pt}{0.400pt}}
\put(1028.0,982.0){\rule[-0.200pt]{0.400pt}{4.818pt}}
\put(1094.0,982.0){\rule[-0.200pt]{0.400pt}{4.818pt}}
\put(944.0,389.0){\rule[-0.200pt]{0.400pt}{8.672pt}}
\put(934.0,389.0){\rule[-0.200pt]{4.818pt}{0.400pt}}
\put(934.0,425.0){\rule[-0.200pt]{4.818pt}{0.400pt}}
\put(476.0,330.0){\rule[-0.200pt]{0.400pt}{13.009pt}}
\put(466.0,330.0){\rule[-0.200pt]{4.818pt}{0.400pt}}
\put(466.0,384.0){\rule[-0.200pt]{4.818pt}{0.400pt}}
\put(944,693){\circle*{24}}
\put(476,612){\circle*{24}}
\put(944.0,675.0){\rule[-0.200pt]{0.400pt}{8.431pt}}
\put(934.0,675.0){\rule[-0.200pt]{4.818pt}{0.400pt}}
\put(934.0,710.0){\rule[-0.200pt]{4.818pt}{0.400pt}}
\put(476.0,592.0){\rule[-0.200pt]{0.400pt}{9.636pt}}
\put(466.0,592.0){\rule[-0.200pt]{4.818pt}{0.400pt}}
\put(466.0,632.0){\rule[-0.200pt]{4.818pt}{0.400pt}}
\put(1006,947){\makebox(0,0)[r]{{\small plaq SW}}}
\put(1050,947){\circle{12}}
\put(274,377){\circle{12}}
\put(304,331){\circle{12}}
\put(350,360){\circle{12}}
\put(439,374){\circle{12}}
\put(627,415){\circle{12}}
\put(877,461){\circle{12}}
\put(1028.0,947.0){\rule[-0.200pt]{15.899pt}{0.400pt}}
\put(1028.0,937.0){\rule[-0.200pt]{0.400pt}{4.818pt}}
\put(1094.0,937.0){\rule[-0.200pt]{0.400pt}{4.818pt}}
\put(274.0,362.0){\rule[-0.200pt]{0.400pt}{7.227pt}}
\put(264.0,362.0){\rule[-0.200pt]{4.818pt}{0.400pt}}
\put(264.0,392.0){\rule[-0.200pt]{4.818pt}{0.400pt}}
\put(304.0,316.0){\rule[-0.200pt]{0.400pt}{6.986pt}}
\put(294.0,316.0){\rule[-0.200pt]{4.818pt}{0.400pt}}
\put(294.0,345.0){\rule[-0.200pt]{4.818pt}{0.400pt}}
\put(350.0,345.0){\rule[-0.200pt]{0.400pt}{7.468pt}}
\put(340.0,345.0){\rule[-0.200pt]{4.818pt}{0.400pt}}
\put(340.0,376.0){\rule[-0.200pt]{4.818pt}{0.400pt}}
\put(439.0,368.0){\rule[-0.200pt]{0.400pt}{3.132pt}}
\put(429.0,368.0){\rule[-0.200pt]{4.818pt}{0.400pt}}
\put(429.0,381.0){\rule[-0.200pt]{4.818pt}{0.400pt}}
\put(627.0,408.0){\rule[-0.200pt]{0.400pt}{3.373pt}}
\put(617.0,408.0){\rule[-0.200pt]{4.818pt}{0.400pt}}
\put(617.0,422.0){\rule[-0.200pt]{4.818pt}{0.400pt}}
\put(877.0,453.0){\rule[-0.200pt]{0.400pt}{3.613pt}}
\put(867.0,453.0){\rule[-0.200pt]{4.818pt}{0.400pt}}
\put(867.0,468.0){\rule[-0.200pt]{4.818pt}{0.400pt}}
\put(274,685){\circle{12}}
\put(304,567){\circle{12}}
\put(350,622){\circle{12}}
\put(439,607){\circle{12}}
\put(627,650){\circle{12}}
\put(877,678){\circle{12}}
\put(274.0,625.0){\rule[-0.200pt]{0.400pt}{28.908pt}}
\put(264.0,625.0){\rule[-0.200pt]{4.818pt}{0.400pt}}
\put(264.0,745.0){\rule[-0.200pt]{4.818pt}{0.400pt}}
\put(304.0,537.0){\rule[-0.200pt]{0.400pt}{14.695pt}}
\put(294.0,537.0){\rule[-0.200pt]{4.818pt}{0.400pt}}
\put(294.0,598.0){\rule[-0.200pt]{4.818pt}{0.400pt}}
\put(350.0,606.0){\rule[-0.200pt]{0.400pt}{7.468pt}}
\put(340.0,606.0){\rule[-0.200pt]{4.818pt}{0.400pt}}
\put(340.0,637.0){\rule[-0.200pt]{4.818pt}{0.400pt}}
\put(439.0,597.0){\rule[-0.200pt]{0.400pt}{4.818pt}}
\put(429.0,597.0){\rule[-0.200pt]{4.818pt}{0.400pt}}
\put(429.0,617.0){\rule[-0.200pt]{4.818pt}{0.400pt}}
\put(627.0,642.0){\rule[-0.200pt]{0.400pt}{4.095pt}}
\put(617.0,642.0){\rule[-0.200pt]{4.818pt}{0.400pt}}
\put(617.0,659.0){\rule[-0.200pt]{4.818pt}{0.400pt}}
\put(877.0,668.0){\rule[-0.200pt]{0.400pt}{5.059pt}}
\put(867.0,668.0){\rule[-0.200pt]{4.818pt}{0.400pt}}
\put(867.0,689.0){\rule[-0.200pt]{4.818pt}{0.400pt}}
\sbox{\plotpoint}{\rule[-0.500pt]{1.000pt}{1.000pt}}%
\put(176,327){\usebox{\plotpoint}}
\put(176.00,327.00){\usebox{\plotpoint}}
\multiput(186,329)(20.261,4.503){0}{\usebox{\plotpoint}}
\put(196.31,331.26){\usebox{\plotpoint}}
\multiput(205,333)(20.352,4.070){0}{\usebox{\plotpoint}}
\put(216.66,335.37){\usebox{\plotpoint}}
\multiput(224,337)(20.652,2.065){0}{\usebox{\plotpoint}}
\put(237.12,338.62){\usebox{\plotpoint}}
\multiput(244,340)(20.352,4.070){0}{\usebox{\plotpoint}}
\put(257.46,342.77){\usebox{\plotpoint}}
\multiput(263,344)(20.352,4.070){0}{\usebox{\plotpoint}}
\put(277.79,346.96){\usebox{\plotpoint}}
\multiput(283,348)(20.261,4.503){0}{\usebox{\plotpoint}}
\put(298.19,350.62){\usebox{\plotpoint}}
\multiput(302,351)(20.352,4.070){0}{\usebox{\plotpoint}}
\put(318.57,354.46){\usebox{\plotpoint}}
\multiput(321,355)(20.352,4.070){0}{\usebox{\plotpoint}}
\put(338.91,358.58){\usebox{\plotpoint}}
\multiput(341,359)(20.352,4.070){0}{\usebox{\plotpoint}}
\put(359.37,361.93){\usebox{\plotpoint}}
\multiput(360,362)(20.352,4.070){0}{\usebox{\plotpoint}}
\put(379.73,365.95){\usebox{\plotpoint}}
\multiput(380,366)(20.261,4.503){0}{\usebox{\plotpoint}}
\multiput(389,368)(20.352,4.070){0}{\usebox{\plotpoint}}
\put(400.04,370.21){\usebox{\plotpoint}}
\multiput(409,372)(20.629,2.292){0}{\usebox{\plotpoint}}
\put(420.52,373.50){\usebox{\plotpoint}}
\multiput(428,375)(20.352,4.070){0}{\usebox{\plotpoint}}
\put(440.87,377.57){\usebox{\plotpoint}}
\multiput(448,379)(20.261,4.503){0}{\usebox{\plotpoint}}
\put(461.18,381.84){\usebox{\plotpoint}}
\multiput(467,383)(20.652,2.065){0}{\usebox{\plotpoint}}
\put(481.66,385.04){\usebox{\plotpoint}}
\multiput(486,386)(20.352,4.070){0}{\usebox{\plotpoint}}
\put(501.99,389.20){\usebox{\plotpoint}}
\multiput(506,390)(20.261,4.503){0}{\usebox{\plotpoint}}
\put(522.30,393.46){\usebox{\plotpoint}}
\multiput(525,394)(20.352,4.070){0}{\usebox{\plotpoint}}
\put(542.76,396.86){\usebox{\plotpoint}}
\multiput(544,397)(20.352,4.070){0}{\usebox{\plotpoint}}
\put(563.13,400.83){\usebox{\plotpoint}}
\multiput(564,401)(20.352,4.070){0}{\usebox{\plotpoint}}
\multiput(574,403)(20.261,4.503){0}{\usebox{\plotpoint}}
\put(583.44,405.09){\usebox{\plotpoint}}
\multiput(593,407)(20.652,2.065){0}{\usebox{\plotpoint}}
\put(603.93,408.21){\usebox{\plotpoint}}
\multiput(612,410)(20.352,4.070){0}{\usebox{\plotpoint}}
\put(624.25,412.45){\usebox{\plotpoint}}
\multiput(632,414)(20.261,4.503){0}{\usebox{\plotpoint}}
\put(644.56,416.71){\usebox{\plotpoint}}
\multiput(651,418)(20.652,2.065){0}{\usebox{\plotpoint}}
\put(665.06,419.81){\usebox{\plotpoint}}
\multiput(671,421)(20.261,4.503){0}{\usebox{\plotpoint}}
\put(685.37,424.07){\usebox{\plotpoint}}
\multiput(690,425)(20.352,4.070){0}{\usebox{\plotpoint}}
\put(705.70,428.27){\usebox{\plotpoint}}
\multiput(709,429)(20.652,2.065){0}{\usebox{\plotpoint}}
\put(726.18,431.44){\usebox{\plotpoint}}
\multiput(729,432)(20.261,4.503){0}{\usebox{\plotpoint}}
\put(746.49,435.70){\usebox{\plotpoint}}
\multiput(748,436)(20.352,4.070){0}{\usebox{\plotpoint}}
\put(766.85,439.77){\usebox{\plotpoint}}
\multiput(768,440)(20.261,4.503){0}{\usebox{\plotpoint}}
\multiput(777,442)(20.652,2.065){0}{\usebox{\plotpoint}}
\put(787.30,443.06){\usebox{\plotpoint}}
\multiput(797,445)(20.261,4.503){0}{\usebox{\plotpoint}}
\put(807.62,447.32){\usebox{\plotpoint}}
\multiput(816,449)(20.352,4.070){0}{\usebox{\plotpoint}}
\put(827.96,451.44){\usebox{\plotpoint}}
\multiput(835,453)(20.652,2.065){0}{\usebox{\plotpoint}}
\put(848.43,454.69){\usebox{\plotpoint}}
\multiput(855,456)(20.261,4.503){0}{\usebox{\plotpoint}}
\put(868.74,458.95){\usebox{\plotpoint}}
\multiput(874,460)(20.352,4.070){0}{\usebox{\plotpoint}}
\put(889.09,463.02){\usebox{\plotpoint}}
\multiput(894,464)(20.629,2.292){0}{\usebox{\plotpoint}}
\put(909.56,466.31){\usebox{\plotpoint}}
\multiput(913,467)(20.352,4.070){0}{\usebox{\plotpoint}}
\put(929.88,470.53){\usebox{\plotpoint}}
\multiput(932,471)(20.352,4.070){0}{\usebox{\plotpoint}}
\put(950.23,474.65){\usebox{\plotpoint}}
\multiput(952,475)(20.629,2.292){0}{\usebox{\plotpoint}}
\put(970.70,477.94){\usebox{\plotpoint}}
\multiput(971,478)(20.352,4.070){0}{\usebox{\plotpoint}}
\multiput(981,480)(20.352,4.070){0}{\usebox{\plotpoint}}
\put(991.05,482.01){\usebox{\plotpoint}}
\multiput(1000,484)(20.352,4.070){0}{\usebox{\plotpoint}}
\put(1011.36,486.27){\usebox{\plotpoint}}
\multiput(1020,488)(20.629,2.292){0}{\usebox{\plotpoint}}
\put(1031.84,489.57){\usebox{\plotpoint}}
\multiput(1039,491)(20.352,4.070){0}{\usebox{\plotpoint}}
\put(1052.18,493.71){\usebox{\plotpoint}}
\multiput(1058,495)(20.352,4.070){0}{\usebox{\plotpoint}}
\put(1072.50,497.90){\usebox{\plotpoint}}
\multiput(1078,499)(20.652,2.065){0}{\usebox{\plotpoint}}
\put(1092.98,501.11){\usebox{\plotpoint}}
\multiput(1097,502)(20.352,4.070){0}{\usebox{\plotpoint}}
\put(1113.31,505.26){\usebox{\plotpoint}}
\multiput(1117,506)(20.261,4.503){0}{\usebox{\plotpoint}}
\put(1133.62,509.52){\usebox{\plotpoint}}
\put(1136,510){\usebox{\plotpoint}}
\put(176,581){\usebox{\plotpoint}}
\put(176.00,581.00){\usebox{\plotpoint}}
\multiput(186,583)(20.629,2.292){0}{\usebox{\plotpoint}}
\put(196.49,584.15){\usebox{\plotpoint}}
\multiput(205,585)(20.352,4.070){0}{\usebox{\plotpoint}}
\put(217.00,587.22){\usebox{\plotpoint}}
\multiput(224,588)(20.652,2.065){0}{\usebox{\plotpoint}}
\put(237.59,589.72){\usebox{\plotpoint}}
\multiput(244,591)(20.652,2.065){0}{\usebox{\plotpoint}}
\put(258.14,592.46){\usebox{\plotpoint}}
\multiput(263,593)(20.352,4.070){0}{\usebox{\plotpoint}}
\put(278.64,595.56){\usebox{\plotpoint}}
\multiput(283,596)(20.261,4.503){0}{\usebox{\plotpoint}}
\put(299.12,598.71){\usebox{\plotpoint}}
\multiput(302,599)(20.652,2.065){0}{\usebox{\plotpoint}}
\put(319.63,601.69){\usebox{\plotpoint}}
\multiput(321,602)(20.652,2.065){0}{\usebox{\plotpoint}}
\put(340.25,603.93){\usebox{\plotpoint}}
\multiput(341,604)(20.352,4.070){0}{\usebox{\plotpoint}}
\multiput(351,606)(20.629,2.292){0}{\usebox{\plotpoint}}
\put(360.74,607.15){\usebox{\plotpoint}}
\multiput(370,609)(20.652,2.065){0}{\usebox{\plotpoint}}
\put(381.25,610.14){\usebox{\plotpoint}}
\multiput(389,611)(20.352,4.070){0}{\usebox{\plotpoint}}
\put(401.75,613.27){\usebox{\plotpoint}}
\multiput(409,614)(20.629,2.292){0}{\usebox{\plotpoint}}
\put(422.32,615.86){\usebox{\plotpoint}}
\multiput(428,617)(20.652,2.065){0}{\usebox{\plotpoint}}
\put(442.89,618.49){\usebox{\plotpoint}}
\multiput(448,619)(20.261,4.503){0}{\usebox{\plotpoint}}
\put(463.37,621.64){\usebox{\plotpoint}}
\multiput(467,622)(20.352,4.070){0}{\usebox{\plotpoint}}
\put(483.87,624.76){\usebox{\plotpoint}}
\multiput(486,625)(20.652,2.065){0}{\usebox{\plotpoint}}
\put(504.39,627.68){\usebox{\plotpoint}}
\multiput(506,628)(20.629,2.292){0}{\usebox{\plotpoint}}
\multiput(515,629)(20.652,2.065){0}{\usebox{\plotpoint}}
\put(525.01,630.00){\usebox{\plotpoint}}
\multiput(535,632)(20.629,2.292){0}{\usebox{\plotpoint}}
\put(545.49,633.30){\usebox{\plotpoint}}
\multiput(554,635)(20.652,2.065){0}{\usebox{\plotpoint}}
\put(566.01,636.20){\usebox{\plotpoint}}
\multiput(574,637)(20.261,4.503){0}{\usebox{\plotpoint}}
\put(586.49,639.35){\usebox{\plotpoint}}
\multiput(593,640)(20.652,2.065){0}{\usebox{\plotpoint}}
\put(607.07,641.90){\usebox{\plotpoint}}
\multiput(612,643)(20.652,2.065){0}{\usebox{\plotpoint}}
\put(627.62,644.56){\usebox{\plotpoint}}
\multiput(632,645)(20.261,4.503){0}{\usebox{\plotpoint}}
\put(648.10,647.71){\usebox{\plotpoint}}
\multiput(651,648)(20.352,4.070){0}{\usebox{\plotpoint}}
\put(668.61,650.76){\usebox{\plotpoint}}
\multiput(671,651)(20.629,2.292){0}{\usebox{\plotpoint}}
\put(689.11,653.82){\usebox{\plotpoint}}
\multiput(690,654)(20.652,2.065){0}{\usebox{\plotpoint}}
\multiput(700,655)(20.629,2.292){0}{\usebox{\plotpoint}}
\put(709.73,656.15){\usebox{\plotpoint}}
\multiput(719,658)(20.652,2.065){0}{\usebox{\plotpoint}}
\put(730.25,659.14){\usebox{\plotpoint}}
\multiput(738,660)(20.352,4.070){0}{\usebox{\plotpoint}}
\put(750.74,662.27){\usebox{\plotpoint}}
\multiput(758,663)(20.352,4.070){0}{\usebox{\plotpoint}}
\put(771.24,665.36){\usebox{\plotpoint}}
\multiput(777,666)(20.652,2.065){0}{\usebox{\plotpoint}}
\put(791.82,667.96){\usebox{\plotpoint}}
\multiput(797,669)(20.629,2.292){0}{\usebox{\plotpoint}}
\put(812.38,670.64){\usebox{\plotpoint}}
\multiput(816,671)(20.352,4.070){0}{\usebox{\plotpoint}}
\put(832.88,673.76){\usebox{\plotpoint}}
\multiput(835,674)(20.352,4.070){0}{\usebox{\plotpoint}}
\put(853.38,676.84){\usebox{\plotpoint}}
\multiput(855,677)(20.629,2.292){0}{\usebox{\plotpoint}}
\put(873.88,679.98){\usebox{\plotpoint}}
\multiput(874,680)(20.652,2.065){0}{\usebox{\plotpoint}}
\multiput(884,681)(20.652,2.065){0}{\usebox{\plotpoint}}
\put(894.52,682.12){\usebox{\plotpoint}}
\multiput(903,684)(20.652,2.065){0}{\usebox{\plotpoint}}
\put(915.01,685.20){\usebox{\plotpoint}}
\multiput(923,686)(20.261,4.503){0}{\usebox{\plotpoint}}
\put(935.49,688.35){\usebox{\plotpoint}}
\multiput(942,689)(20.352,4.070){0}{\usebox{\plotpoint}}
\put(955.99,691.44){\usebox{\plotpoint}}
\multiput(961,692)(20.652,2.065){0}{\usebox{\plotpoint}}
\put(976.55,694.11){\usebox{\plotpoint}}
\multiput(981,695)(20.652,2.065){0}{\usebox{\plotpoint}}
\put(997.13,696.68){\usebox{\plotpoint}}
\multiput(1000,697)(20.352,4.070){0}{\usebox{\plotpoint}}
\put(1017.64,699.76){\usebox{\plotpoint}}
\multiput(1020,700)(20.261,4.503){0}{\usebox{\plotpoint}}
\put(1038.11,702.91){\usebox{\plotpoint}}
\multiput(1039,703)(20.652,2.065){0}{\usebox{\plotpoint}}
\multiput(1049,704)(20.261,4.503){0}{\usebox{\plotpoint}}
\put(1058.59,706.06){\usebox{\plotpoint}}
\multiput(1068,707)(20.652,2.065){0}{\usebox{\plotpoint}}
\put(1079.23,708.25){\usebox{\plotpoint}}
\multiput(1088,710)(20.629,2.292){0}{\usebox{\plotpoint}}
\put(1099.74,711.27){\usebox{\plotpoint}}
\multiput(1107,712)(20.352,4.070){0}{\usebox{\plotpoint}}
\put(1120.24,714.36){\usebox{\plotpoint}}
\multiput(1126,715)(20.352,4.070){0}{\usebox{\plotpoint}}
\put(1136,717){\usebox{\plotpoint}}
\end{picture}
\end{center}
\caption{ 
$D/V$ and $N/V$ for D234c and plaq TI SW
(SW data kindly supplied by SCRI.)
}
\label{fig:ratios}
\end{figure}

We also examined baryon masses; ratios of these to the vector mass
are shown in Fig.~\ref{fig:ratios}. Our measured values at $a=0.25~\fm$
are within $2\si$ of the quadratic extrapolation of the SCRI values
to the continuum.

\def\st{\rule[-1.5ex]{0em}{4ex}} 
\begin{table}[htb]
\begin{tabular}{lllll}
\hline
\st $a$ & mean-link D234c & mean-link SW & plaquette SW & plaquette Wilson\\
\hline
 0.4  & 0.969(9)  & 0.933(10) &  \\
 0.25 & 1.027(10) & 1.041(15) &  \\
 0    &            &            & 1.035(5) & 1.034(37) \\
\hline
\end{tabular}
\caption{
Phi (V) masses in GeV at $P/V = 0.70$, from 
tables \protect\ref{tab:glue} and \protect\ref{tab:hadrons70}, 
using a quadratic
fit to SCRI's SW data to extrapolate it to $a=0$.
For the D234c action we see a $2(1)\%$ scaling error at
$a=0.25~\fm$, and $7(1)\%$ scaling error at $a=0.4~\fm$.
}
\label{tab:phi70GeV}
\end{table}

In table \ref{tab:phi70GeV} we give the vector mass at our two 
lattice spacings, along with a naive continuum extrapolation
of SCRI's data.
(For details of the scale determination see \ref{app:lattspacing}).
The errors include the relative error between our mass scale
and SCRI's, but not the uncertainty in the absolute scale, which
may be affected by quenching effects.

Using a quadratic
fit to SCRI's SW data to extrapolate it to $a=0$, 
the D234c action shows a $2(1)\%$ finite-$a$ error at
$a=0.25~\fm$, and $7(1)\%$ finite-$a$ error at $a=0.4~\fm$.
(The quadratic fit may be too naive: the true continuum value
could differ by a few percent, however this will not substantially affect
our conclusions below.)
These finite-$a$ errors are due to  radiative
corrections to the tree-level coupling constants, and higher-order
interactions not included in our action. We measured the sensitivity of the
hadron masses to radiative corrections by varying each of the
tree-level coupling 
constants. The fractional change caused by multiplying
each coupling constant in turn by $1+\alpha_s$ is shown in
Table~\ref{tab:sensitivity}. The only coupling for which radiative
corrections are important is the clover coupling,~$C_F$ \cite{Tim}.
A perturbative analysis of
$\Csw$ through $\O(\al_s)$ will soon be completed \cite{TrotLep}.
Alternatively, the $\O(\al_s)$ coefficient could be
determined by making the vector meson mass at our smallest lattice spacing
agree with the continuum.
For example, using
our continuum extrapolation of SCRI's data we find that taking
\beq\label{CF}
C_F = 1 + 0.2\alpha_s,
\eeq
where $\al_s$ is the bare coupling \eqn{alpha}, reduces the $V$ mass
error at 0.25~\fm\ from 2(1)\% to 0(1)\%, and at 0.4~\fm\ from 7(1)\%
to 0(1)\%.  This suggests that perturbative corrections to $\Csw$ are
relatively small after tadpole improvement.  The necessary corrections
to $\Csw$ are still perturbative if the true continuum value differs
by a few percent from the naive extrapolation of the SCRI data.

\def\phXXX{\protect\phantom{XXX}}
\def\phMinus{\protect\phantom{-}}
\def\st{\rule[-1.5ex]{0em}{4ex}} 
\begin{table} 
\begin{tabular}{l@{\phXXX}ll@{\phXXX}
l@{\phXXX}l@{\phXXX}l}
\hline
\st coeff & \multicolumn{2}{c}{$\Csw$} & $C_3$ & $\phMinus C_4$ & $\phMinus C_{2F}$ \\
\st $a$ (fm) & 0.25 & 0.4        & 0.4   & $\phMinus 0.4$   & $\phMinus 0.4$ \\
\hline
\st P & 11(1)\% & 35(1)\%  & 3.8(6)\% & $-1.0(2)\%$ & $-2.9(9)\%$  \\
\st V & 10(1)\% & 35(1)\%  & 3.7(4)\% & $-1.0(2)\%$ & $-2.9(5)\%$  \\
\st N & 6(2)\%  & 26(2)\%  & 3.4(7)\% & $-0.9(2)\%$ & $-1.7(7)\%$  \\
\st D & 8(1)\%  & 25(2)\%  & 3.8(9)\% & $-1.1(3)\%$ & $-2.3(4)\% $ \\
\hline
\end{tabular}
\caption{
Percentage change in hadron masses 
when individual coefficients in the mean-link tree-level TI
D234c quark action (Eq.~\protect\ref{action}) are multiplied by $1+\alpha_s$.
All are at $P/V\approx 0.7$, except the $\Csw, a=0.4$ is at $P/V=0.76$.}
\label{tab:sensitivity}
\end{table}

We also include SW results in Figure~\ref{fig:phi70} using
the Landau-link tadpole-improved tree-level value for the clover
coefficient,~$\Csw$. Comparing these
with SCRI's SW results, for which the plaquette was used to determine~$u_0$,
we see that the Landau-link results show much smaller
finite-$a$ errors.
This result was anticipated based
on work at smaller lattice spacings using SW quarks and the
(unimproved) Wilson action for the gluons\,\cite{gplTsukuba}.

\hide{
It is surprising that the SW and D234c results, both with Landau-link
tadpole-improvement, give results that are quite close even at
0.4\,fm. This is also apparent in Table~\ref{tab:sensitivity} where we show
that the hadron masses are almost insensitive to the value of the
coupling constant $C_3$, which multiplies the leading correction term
beyond the SW formalism. The differences we see between SW and D234c
are mostly due to order~$a^3$ terms. 
}

\subsection{Hadron dispersion relations}

\begin{figure}
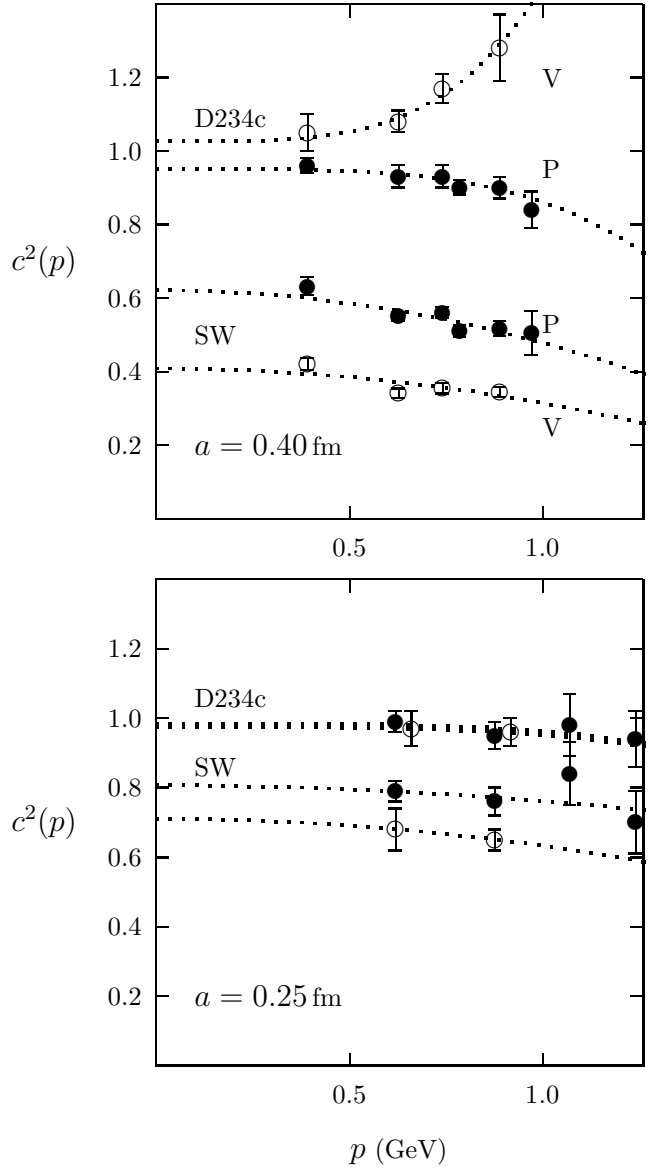
 
\begin{center}
\setlength{\unitlength}{0.240900pt}
\ifx\plotpoint\undefined\newsavebox{\plotpoint}\fi
\sbox{\plotpoint}{\rule[-0.200pt]{0.400pt}{0.400pt}}%


\end{center}
\caption{ 
Speed of light squared using the D234c and SW actions, for the
pseudoscalar ($\bullet$) and vector ($\circ$) mesons, 
at $P/V=0.76$, on a $5^2\times 8 \times 18$
lattice at $a=0.4~\fm$ (see Table \protect\ref{tab:c2_b}), and on a
$8^3\times 24$ lattice at $a=0.25$\,fm (see Table \protect\ref{tab:c2_a0.25}).
[D234c V points are offset for clarity.]
D234c shows much better rotational invariance; both discretizations improve
at the smaller lattice spacing, where the pseudoscalar and vector
results for D234c are indistinguishable with our statistics.}
\label{fig:csq}
\end{figure}

The finite-$a$ errors in the $\phi$ mass appear to be almost as
small for the SW action as they are for D234c,
but this is deceptive. To see why, we consider the quantity
\beq
c^2(\pv) = \frac{E^2(\pv)-E^2(0)}{\pv^2}
\eeq
for different hadrons and three momenta~$\pv$, where $E(\pv)$ is the
hadron's energy. In the continuum limit, $c^2(\pv)=1$ for all $\pv$.
This quantity is particularly sensistive to the $C_3$~term
in the D234c action since this term is not rotationally invariant; it
cancels the leading (non-rotationally invariant) error in the SW
action. Our results for $c^2$, for both pseudoscalar and vector
mesons, are shown in Figure~\ref{fig:csq}. At 0.4\,fm, D234c is
dramatically superior: it deviates from $c^2\seq1$ by only 3--5\% at
zero momentum, and by less than 10\% even at momenta of order $1.5/a$,
while SW gives results that deviate by 40--60\% or more for all
momenta, including zero. As expected both formalisms improve at
0.25\,fm, although D234c is still clearly superior.

The $c^2$ results have practical implications. 
For example, there are two different definitions of a hadron's mass in
lattice simulations. One is the ``static mass,'' 
\beq
m_{\rm static} \equiv E(p\seq0),
\eeq
and the other the ``kinetic mass,'' 
\beq
m_{\rm kinetic} \equiv \lim_{p\to0} \frac{p}{dE/dp} = E(0)/c^2(0).
\eeq
In D234c these two
definitions agree to within~3--5\% at~0.4\,fm. In SW they differ by 40--60\%,
making it impossible to say what the ``true'' mass is for this
formalism at~0.4\,fm. The
kinetic and static masses in either formalism must be equal for zero-mass
mesons, because of the axis-interchange symmetry of the actions. The
deviations seen here are because the strange quark is relatively
massive at our lattice spacings:
the $\phi$~mass, for example, is~$2.1/a$ with D234c at~0.4\,fm.
These results illustrate that D234c is far more accurate than SW for hadrons
with large masses in lattice units. 
More generally, D234c is far more accurate for hadrons
with large energies and/or large momenta. 

\begin{figure} 
\begin{center}
\setlength{\unitlength}{0.240900pt}
\ifx\plotpoint\undefined\newsavebox{\plotpoint}\fi
\sbox{\plotpoint}{\rule[-0.200pt]{0.400pt}{0.400pt}}%
\begin{picture}(1200,1080)(0,0)
\font\gnuplot=cmr10 at 10pt
\gnuplot
\sbox{\plotpoint}{\rule[-0.200pt]{0.400pt}{0.400pt}}%
\put(220.0,113.0){\rule[-0.200pt]{0.400pt}{227.410pt}}
\put(220.0,258.0){\rule[-0.200pt]{4.818pt}{0.400pt}}
\put(198,258){\makebox(0,0)[r]{0.8}}
\put(1116.0,258.0){\rule[-0.200pt]{4.818pt}{0.400pt}}
\put(220.0,549.0){\rule[-0.200pt]{4.818pt}{0.400pt}}
\put(198,549){\makebox(0,0)[r]{1.0}}
\put(1116.0,549.0){\rule[-0.200pt]{4.818pt}{0.400pt}}
\put(220.0,839.0){\rule[-0.200pt]{4.818pt}{0.400pt}}
\put(198,839){\makebox(0,0)[r]{1.2}}
\put(1116.0,839.0){\rule[-0.200pt]{4.818pt}{0.400pt}}
\put(602.0,113.0){\rule[-0.200pt]{0.400pt}{4.818pt}}
\put(602,68){\makebox(0,0){0.5}}
\put(602.0,1037.0){\rule[-0.200pt]{0.400pt}{4.818pt}}
\put(983.0,113.0){\rule[-0.200pt]{0.400pt}{4.818pt}}
\put(983,68){\makebox(0,0){1.0}}
\put(983.0,1037.0){\rule[-0.200pt]{0.400pt}{4.818pt}}
\put(220.0,113.0){\rule[-0.200pt]{220.664pt}{0.400pt}}
\put(1136.0,113.0){\rule[-0.200pt]{0.400pt}{227.410pt}}
\put(220.0,1057.0){\rule[-0.200pt]{220.664pt}{0.400pt}}
\put(45,585){\makebox(0,0){$c^2(p)$}}
\put(678,-22){\makebox(0,0){$p$ (GeV)}}
\put(602,440){\makebox(0,0)[l]{P}}
\put(602,520){\makebox(0,0)[l]{P}}
\put(602,694){\makebox(0,0)[l]{V}}
\put(602,941){\makebox(0,0)[l]{V}}
\put(220.0,113.0){\rule[-0.200pt]{0.400pt}{227.410pt}}
\put(525,331){\makebox(0,0)[r]{$C_3=1.2$, P}}
\put(569,331){\circle*{24}}
\put(519,563){\circle*{24}}
\put(699,563){\circle*{24}}
\put(786,578){\circle*{24}}
\put(819,534){\circle*{24}}
\put(899,665){\circle*{24}}
\put(962,563){\circle*{24}}
\put(547.0,331.0){\rule[-0.200pt]{15.899pt}{0.400pt}}
\put(547.0,321.0){\rule[-0.200pt]{0.400pt}{4.818pt}}
\put(613.0,321.0){\rule[-0.200pt]{0.400pt}{4.818pt}}
\put(519.0,534.0){\rule[-0.200pt]{0.400pt}{13.972pt}}
\put(509.0,534.0){\rule[-0.200pt]{4.818pt}{0.400pt}}
\put(509.0,592.0){\rule[-0.200pt]{4.818pt}{0.400pt}}
\put(699.0,534.0){\rule[-0.200pt]{0.400pt}{13.972pt}}
\put(689.0,534.0){\rule[-0.200pt]{4.818pt}{0.400pt}}
\put(689.0,592.0){\rule[-0.200pt]{4.818pt}{0.400pt}}
\put(786.0,549.0){\rule[-0.200pt]{0.400pt}{13.972pt}}
\put(776.0,549.0){\rule[-0.200pt]{4.818pt}{0.400pt}}
\put(776.0,607.0){\rule[-0.200pt]{4.818pt}{0.400pt}}
\put(819.0,505.0){\rule[-0.200pt]{0.400pt}{13.972pt}}
\put(809.0,505.0){\rule[-0.200pt]{4.818pt}{0.400pt}}
\put(809.0,563.0){\rule[-0.200pt]{4.818pt}{0.400pt}}
\put(899.0,621.0){\rule[-0.200pt]{0.400pt}{20.958pt}}
\put(889.0,621.0){\rule[-0.200pt]{4.818pt}{0.400pt}}
\put(889.0,708.0){\rule[-0.200pt]{4.818pt}{0.400pt}}
\put(962.0,447.0){\rule[-0.200pt]{0.400pt}{55.889pt}}
\put(952.0,447.0){\rule[-0.200pt]{4.818pt}{0.400pt}}
\put(952.0,679.0){\rule[-0.200pt]{4.818pt}{0.400pt}}
\put(525,286){\makebox(0,0)[r]{$C_3=1.2$, V}}
\put(569,286){\circle{24}}
\put(519,883){\circle{24}}
\put(699,883){\circle{24}}
\put(786,1013){\circle{24}}
\put(547.0,286.0){\rule[-0.200pt]{15.899pt}{0.400pt}}
\put(547.0,276.0){\rule[-0.200pt]{0.400pt}{4.818pt}}
\put(613.0,276.0){\rule[-0.200pt]{0.400pt}{4.818pt}}
\put(519.0,810.0){\rule[-0.200pt]{0.400pt}{34.930pt}}
\put(509.0,810.0){\rule[-0.200pt]{4.818pt}{0.400pt}}
\put(509.0,955.0){\rule[-0.200pt]{4.818pt}{0.400pt}}
\put(699.0,796.0){\rule[-0.200pt]{0.400pt}{41.917pt}}
\put(689.0,796.0){\rule[-0.200pt]{4.818pt}{0.400pt}}
\put(689.0,970.0){\rule[-0.200pt]{4.818pt}{0.400pt}}
\put(786.0,926.0){\rule[-0.200pt]{0.400pt}{31.558pt}}
\put(776.0,926.0){\rule[-0.200pt]{4.818pt}{0.400pt}}
\put(776.0,1057.0){\rule[-0.200pt]{4.818pt}{0.400pt}}
\sbox{\plotpoint}{\rule[-0.500pt]{1.000pt}{1.000pt}}%
\put(220,857){\usebox{\plotpoint}}
\put(220.00,857.00){\usebox{\plotpoint}}
\multiput(229,857)(20.756,0.000){0}{\usebox{\plotpoint}}
\put(240.76,857.00){\usebox{\plotpoint}}
\multiput(248,857)(20.756,0.000){0}{\usebox{\plotpoint}}
\put(261.51,857.00){\usebox{\plotpoint}}
\multiput(266,857)(20.756,0.000){0}{\usebox{\plotpoint}}
\put(282.27,857.00){\usebox{\plotpoint}}
\multiput(285,857)(20.756,0.000){0}{\usebox{\plotpoint}}
\multiput(294,857)(20.756,0.000){0}{\usebox{\plotpoint}}
\put(303.02,857.00){\usebox{\plotpoint}}
\multiput(313,857)(20.756,0.000){0}{\usebox{\plotpoint}}
\put(323.78,857.00){\usebox{\plotpoint}}
\multiput(331,857)(20.756,0.000){0}{\usebox{\plotpoint}}
\put(344.53,857.00){\usebox{\plotpoint}}
\multiput(350,857)(20.756,0.000){0}{\usebox{\plotpoint}}
\put(365.29,857.00){\usebox{\plotpoint}}
\multiput(368,857)(20.756,0.000){0}{\usebox{\plotpoint}}
\put(386.00,857.90){\usebox{\plotpoint}}
\multiput(387,858)(20.756,0.000){0}{\usebox{\plotpoint}}
\multiput(396,858)(20.756,0.000){0}{\usebox{\plotpoint}}
\put(406.75,858.00){\usebox{\plotpoint}}
\multiput(414,858)(20.652,2.065){0}{\usebox{\plotpoint}}
\put(427.46,859.00){\usebox{\plotpoint}}
\multiput(433,859)(20.629,2.292){0}{\usebox{\plotpoint}}
\put(448.16,860.00){\usebox{\plotpoint}}
\multiput(451,860)(20.652,2.065){0}{\usebox{\plotpoint}}
\put(468.81,861.87){\usebox{\plotpoint}}
\multiput(470,862)(20.629,2.292){0}{\usebox{\plotpoint}}
\multiput(479,863)(20.756,0.000){0}{\usebox{\plotpoint}}
\put(489.48,863.30){\usebox{\plotpoint}}
\multiput(498,865)(20.629,2.292){0}{\usebox{\plotpoint}}
\put(509.99,866.33){\usebox{\plotpoint}}
\multiput(516,867)(20.629,2.292){0}{\usebox{\plotpoint}}
\put(530.54,869.11){\usebox{\plotpoint}}
\multiput(535,870)(20.629,2.292){0}{\usebox{\plotpoint}}
\put(550.98,872.55){\usebox{\plotpoint}}
\multiput(553,873)(20.261,4.503){0}{\usebox{\plotpoint}}
\put(571.29,876.86){\usebox{\plotpoint}}
\multiput(572,877)(20.261,4.503){0}{\usebox{\plotpoint}}
\multiput(581,879)(19.690,6.563){0}{\usebox{\plotpoint}}
\put(591.29,882.29){\usebox{\plotpoint}}
\multiput(599,884)(19.880,5.964){0}{\usebox{\plotpoint}}
\put(611.29,887.76){\usebox{\plotpoint}}
\multiput(618,890)(19.690,6.563){0}{\usebox{\plotpoint}}
\put(630.84,894.71){\usebox{\plotpoint}}
\multiput(636,897)(19.880,5.964){0}{\usebox{\plotpoint}}
\put(650.26,901.90){\usebox{\plotpoint}}
\multiput(655,904)(18.144,10.080){0}{\usebox{\plotpoint}}
\put(668.82,911.14){\usebox{\plotpoint}}
\multiput(673,913)(18.564,9.282){0}{\usebox{\plotpoint}}
\put(687.37,920.43){\usebox{\plotpoint}}
\multiput(692,923)(18.144,10.080){0}{\usebox{\plotpoint}}
\put(705.30,930.87){\usebox{\plotpoint}}
\multiput(710,934)(17.798,10.679){0}{\usebox{\plotpoint}}
\put(722.87,941.91){\usebox{\plotpoint}}
\multiput(729,946)(16.383,12.743){0}{\usebox{\plotpoint}}
\put(739.56,954.22){\usebox{\plotpoint}}
\put(756.29,966.50){\usebox{\plotpoint}}
\multiput(757,967)(15.513,13.789){0}{\usebox{\plotpoint}}
\put(771.86,980.21){\usebox{\plotpoint}}
\multiput(775,983)(14.676,14.676){0}{\usebox{\plotpoint}}
\put(786.85,994.56){\usebox{\plotpoint}}
\put(801.45,1009.27){\usebox{\plotpoint}}
\multiput(803,1011)(13.885,15.427){0}{\usebox{\plotpoint}}
\put(815.33,1024.70){\usebox{\plotpoint}}
\put(829.26,1040.09){\usebox{\plotpoint}}
\multiput(831,1042)(12.453,16.604){0}{\usebox{\plotpoint}}
\put(841.76,1056.64){\usebox{\plotpoint}}
\put(842,1057){\usebox{\plotpoint}}
\put(220,549){\usebox{\plotpoint}}
\put(220.00,549.00){\usebox{\plotpoint}}
\multiput(229,549)(20.756,0.000){0}{\usebox{\plotpoint}}
\put(240.76,549.00){\usebox{\plotpoint}}
\multiput(248,549)(20.756,0.000){0}{\usebox{\plotpoint}}
\put(261.51,549.00){\usebox{\plotpoint}}
\multiput(266,549)(20.756,0.000){0}{\usebox{\plotpoint}}
\put(282.27,549.00){\usebox{\plotpoint}}
\multiput(285,549)(20.756,0.000){0}{\usebox{\plotpoint}}
\multiput(294,549)(20.756,0.000){0}{\usebox{\plotpoint}}
\put(303.02,549.00){\usebox{\plotpoint}}
\multiput(313,549)(20.756,0.000){0}{\usebox{\plotpoint}}
\put(323.78,549.00){\usebox{\plotpoint}}
\multiput(331,549)(20.756,0.000){0}{\usebox{\plotpoint}}
\put(344.53,549.00){\usebox{\plotpoint}}
\multiput(350,549)(20.756,0.000){0}{\usebox{\plotpoint}}
\put(365.29,549.00){\usebox{\plotpoint}}
\multiput(368,549)(20.756,0.000){0}{\usebox{\plotpoint}}
\put(386.04,549.00){\usebox{\plotpoint}}
\multiput(387,549)(20.756,0.000){0}{\usebox{\plotpoint}}
\multiput(396,549)(20.756,0.000){0}{\usebox{\plotpoint}}
\put(406.80,549.00){\usebox{\plotpoint}}
\multiput(414,549)(20.756,0.000){0}{\usebox{\plotpoint}}
\put(427.55,549.00){\usebox{\plotpoint}}
\multiput(433,549)(20.756,0.000){0}{\usebox{\plotpoint}}
\put(448.31,549.00){\usebox{\plotpoint}}
\multiput(451,549)(20.756,0.000){0}{\usebox{\plotpoint}}
\put(469.07,549.00){\usebox{\plotpoint}}
\multiput(470,549)(20.756,0.000){0}{\usebox{\plotpoint}}
\multiput(479,549)(20.629,2.292){0}{\usebox{\plotpoint}}
\put(489.77,550.00){\usebox{\plotpoint}}
\multiput(498,550)(20.756,0.000){0}{\usebox{\plotpoint}}
\put(510.52,550.00){\usebox{\plotpoint}}
\multiput(516,550)(20.756,0.000){0}{\usebox{\plotpoint}}
\put(531.28,550.00){\usebox{\plotpoint}}
\multiput(535,550)(20.629,2.292){0}{\usebox{\plotpoint}}
\put(551.98,551.00){\usebox{\plotpoint}}
\multiput(553,551)(20.756,0.000){0}{\usebox{\plotpoint}}
\multiput(562,551)(20.756,0.000){0}{\usebox{\plotpoint}}
\put(572.73,551.08){\usebox{\plotpoint}}
\multiput(581,552)(20.756,0.000){0}{\usebox{\plotpoint}}
\put(593.43,552.00){\usebox{\plotpoint}}
\multiput(599,552)(20.652,2.065){0}{\usebox{\plotpoint}}
\put(614.14,553.00){\usebox{\plotpoint}}
\multiput(618,553)(20.629,2.292){0}{\usebox{\plotpoint}}
\put(634.84,554.00){\usebox{\plotpoint}}
\multiput(636,554)(20.756,0.000){0}{\usebox{\plotpoint}}
\multiput(646,554)(20.629,2.292){0}{\usebox{\plotpoint}}
\put(655.54,555.06){\usebox{\plotpoint}}
\multiput(664,556)(20.756,0.000){0}{\usebox{\plotpoint}}
\put(676.22,556.32){\usebox{\plotpoint}}
\multiput(683,557)(20.756,0.000){0}{\usebox{\plotpoint}}
\put(696.91,557.55){\usebox{\plotpoint}}
\multiput(701,558)(20.629,2.292){0}{\usebox{\plotpoint}}
\put(717.55,559.76){\usebox{\plotpoint}}
\multiput(720,560)(20.756,0.000){0}{\usebox{\plotpoint}}
\multiput(729,560)(20.629,2.292){0}{\usebox{\plotpoint}}
\put(738.24,561.03){\usebox{\plotpoint}}
\multiput(747,562)(20.652,2.065){0}{\usebox{\plotpoint}}
\put(758.88,563.21){\usebox{\plotpoint}}
\multiput(766,564)(20.629,2.292){0}{\usebox{\plotpoint}}
\put(779.51,565.50){\usebox{\plotpoint}}
\multiput(784,566)(20.352,4.070){0}{\usebox{\plotpoint}}
\put(800.00,568.67){\usebox{\plotpoint}}
\multiput(803,569)(20.629,2.292){0}{\usebox{\plotpoint}}
\put(820.47,571.88){\usebox{\plotpoint}}
\multiput(821,572)(20.652,2.065){0}{\usebox{\plotpoint}}
\multiput(831,573)(20.261,4.503){0}{\usebox{\plotpoint}}
\put(840.94,575.10){\usebox{\plotpoint}}
\multiput(849,576)(20.261,4.503){0}{\usebox{\plotpoint}}
\put(861.36,578.67){\usebox{\plotpoint}}
\multiput(868,580)(20.629,2.292){0}{\usebox{\plotpoint}}
\put(881.81,582.07){\usebox{\plotpoint}}
\multiput(886,583)(20.261,4.503){0}{\usebox{\plotpoint}}
\put(902.11,586.42){\usebox{\plotpoint}}
\multiput(905,587)(20.261,4.503){0}{\usebox{\plotpoint}}
\put(922.14,591.71){\usebox{\plotpoint}}
\multiput(923,592)(20.261,4.503){0}{\usebox{\plotpoint}}
\multiput(932,594)(20.352,4.070){0}{\usebox{\plotpoint}}
\put(942.41,596.14){\usebox{\plotpoint}}
\multiput(951,599)(20.261,4.503){0}{\usebox{\plotpoint}}
\put(962.36,601.79){\usebox{\plotpoint}}
\multiput(969,604)(19.880,5.964){0}{\usebox{\plotpoint}}
\put(982.14,608.05){\usebox{\plotpoint}}
\multiput(988,610)(19.690,6.563){0}{\usebox{\plotpoint}}
\put(1001.83,614.61){\usebox{\plotpoint}}
\multiput(1006,616)(19.880,5.964){0}{\usebox{\plotpoint}}
\put(1021.62,620.87){\usebox{\plotpoint}}
\multiput(1025,622)(18.967,8.430){0}{\usebox{\plotpoint}}
\put(1040.97,628.32){\usebox{\plotpoint}}
\multiput(1043,629)(19.271,7.708){0}{\usebox{\plotpoint}}
\put(1060.17,636.18){\usebox{\plotpoint}}
\multiput(1062,637)(18.967,8.430){0}{\usebox{\plotpoint}}
\put(1079.13,644.61){\usebox{\plotpoint}}
\multiput(1080,645)(19.271,7.708){0}{\usebox{\plotpoint}}
\put(1098.26,652.67){\usebox{\plotpoint}}
\multiput(1099,653)(18.144,10.080){0}{\usebox{\plotpoint}}
\put(1116.82,661.92){\usebox{\plotpoint}}
\multiput(1117,662)(18.564,9.282){0}{\usebox{\plotpoint}}
\put(1135.19,671.55){\usebox{\plotpoint}}
\put(1136,672){\usebox{\plotpoint}}
\end{picture}
\setlength{\unitlength}{0.240900pt}
\ifx\plotpoint\undefined\newsavebox{\plotpoint}\fi
\sbox{\plotpoint}{\rule[-0.200pt]{0.400pt}{0.400pt}}%
\begin{picture}(1200,1080)(0,0)
\font\gnuplot=cmr10 at 10pt
\gnuplot
\sbox{\plotpoint}{\rule[-0.200pt]{0.400pt}{0.400pt}}%
\put(220.0,113.0){\rule[-0.200pt]{0.400pt}{227.410pt}}
\put(220.0,258.0){\rule[-0.200pt]{4.818pt}{0.400pt}}
\put(198,258){\makebox(0,0)[r]{0.8}}
\put(1116.0,258.0){\rule[-0.200pt]{4.818pt}{0.400pt}}
\put(220.0,549.0){\rule[-0.200pt]{4.818pt}{0.400pt}}
\put(198,549){\makebox(0,0)[r]{1.0}}
\put(1116.0,549.0){\rule[-0.200pt]{4.818pt}{0.400pt}}
\put(220.0,839.0){\rule[-0.200pt]{4.818pt}{0.400pt}}
\put(198,839){\makebox(0,0)[r]{1.2}}
\put(1116.0,839.0){\rule[-0.200pt]{4.818pt}{0.400pt}}
\put(602.0,113.0){\rule[-0.200pt]{0.400pt}{4.818pt}}
\put(602,68){\makebox(0,0){0.5}}
\put(602.0,1037.0){\rule[-0.200pt]{0.400pt}{4.818pt}}
\put(983.0,113.0){\rule[-0.200pt]{0.400pt}{4.818pt}}
\put(983,68){\makebox(0,0){1.0}}
\put(983.0,1037.0){\rule[-0.200pt]{0.400pt}{4.818pt}}
\put(220.0,113.0){\rule[-0.200pt]{220.664pt}{0.400pt}}
\put(1136.0,113.0){\rule[-0.200pt]{0.400pt}{227.410pt}}
\put(220.0,1057.0){\rule[-0.200pt]{220.664pt}{0.400pt}}
\put(45,585){\makebox(0,0){$c^2(p)$}}
\put(678,-22){\makebox(0,0){$p$ (GeV)}}
\put(220.0,113.0){\rule[-0.200pt]{0.400pt}{227.410pt}}
\put(525,331){\makebox(0,0)[r]{$C_3=1.2$, P}}
\put(569,331){\circle*{24}}
\put(519,534){\circle*{24}}
\put(699,491){\circle*{24}}
\put(786,578){\circle*{24}}
\put(899,476){\circle*{24}}
\put(962,331){\circle*{24}}
\put(547.0,331.0){\rule[-0.200pt]{15.899pt}{0.400pt}}
\put(547.0,321.0){\rule[-0.200pt]{0.400pt}{4.818pt}}
\put(613.0,321.0){\rule[-0.200pt]{0.400pt}{4.818pt}}
\put(519.0,491.0){\rule[-0.200pt]{0.400pt}{20.958pt}}
\put(509.0,491.0){\rule[-0.200pt]{4.818pt}{0.400pt}}
\put(509.0,578.0){\rule[-0.200pt]{4.818pt}{0.400pt}}
\put(699.0,447.0){\rule[-0.200pt]{0.400pt}{20.958pt}}
\put(689.0,447.0){\rule[-0.200pt]{4.818pt}{0.400pt}}
\put(689.0,534.0){\rule[-0.200pt]{4.818pt}{0.400pt}}
\put(786.0,505.0){\rule[-0.200pt]{0.400pt}{34.930pt}}
\put(776.0,505.0){\rule[-0.200pt]{4.818pt}{0.400pt}}
\put(776.0,650.0){\rule[-0.200pt]{4.818pt}{0.400pt}}
\put(899.0,331.0){\rule[-0.200pt]{0.400pt}{69.861pt}}
\put(889.0,331.0){\rule[-0.200pt]{4.818pt}{0.400pt}}
\put(889.0,621.0){\rule[-0.200pt]{4.818pt}{0.400pt}}
\put(962.0,186.0){\rule[-0.200pt]{0.400pt}{69.861pt}}
\put(952.0,186.0){\rule[-0.200pt]{4.818pt}{0.400pt}}
\put(952.0,476.0){\rule[-0.200pt]{4.818pt}{0.400pt}}
\put(525,286){\makebox(0,0)[r]{$C_3=1.2$, V}}
\put(569,286){\circle{24}}
\put(519,650){\circle{24}}
\put(699,636){\circle{24}}
\put(786,723){\circle{24}}
\put(899,796){\circle{24}}
\put(547.0,286.0){\rule[-0.200pt]{15.899pt}{0.400pt}}
\put(547.0,276.0){\rule[-0.200pt]{0.400pt}{4.818pt}}
\put(613.0,276.0){\rule[-0.200pt]{0.400pt}{4.818pt}}
\put(519.0,578.0){\rule[-0.200pt]{0.400pt}{34.930pt}}
\put(509.0,578.0){\rule[-0.200pt]{4.818pt}{0.400pt}}
\put(509.0,723.0){\rule[-0.200pt]{4.818pt}{0.400pt}}
\put(699.0,592.0){\rule[-0.200pt]{0.400pt}{20.958pt}}
\put(689.0,592.0){\rule[-0.200pt]{4.818pt}{0.400pt}}
\put(689.0,679.0){\rule[-0.200pt]{4.818pt}{0.400pt}}
\put(786.0,679.0){\rule[-0.200pt]{0.400pt}{21.199pt}}
\put(776.0,679.0){\rule[-0.200pt]{4.818pt}{0.400pt}}
\put(776.0,767.0){\rule[-0.200pt]{4.818pt}{0.400pt}}
\put(899.0,708.0){\rule[-0.200pt]{0.400pt}{42.157pt}}
\put(889.0,708.0){\rule[-0.200pt]{4.818pt}{0.400pt}}
\put(889.0,883.0){\rule[-0.200pt]{4.818pt}{0.400pt}}
\sbox{\plotpoint}{\rule[-0.500pt]{1.000pt}{1.000pt}}%
\put(220,537){\usebox{\plotpoint}}
\put(220.00,537.00){\usebox{\plotpoint}}
\multiput(229,537)(20.756,0.000){0}{\usebox{\plotpoint}}
\put(240.76,537.00){\usebox{\plotpoint}}
\multiput(248,537)(20.756,0.000){0}{\usebox{\plotpoint}}
\put(261.51,537.00){\usebox{\plotpoint}}
\multiput(266,537)(20.756,0.000){0}{\usebox{\plotpoint}}
\put(282.27,537.00){\usebox{\plotpoint}}
\multiput(285,537)(20.756,0.000){0}{\usebox{\plotpoint}}
\multiput(294,537)(20.756,0.000){0}{\usebox{\plotpoint}}
\put(303.02,537.00){\usebox{\plotpoint}}
\multiput(313,537)(20.756,0.000){0}{\usebox{\plotpoint}}
\put(323.78,537.00){\usebox{\plotpoint}}
\multiput(331,537)(20.756,0.000){0}{\usebox{\plotpoint}}
\put(344.53,537.00){\usebox{\plotpoint}}
\multiput(350,537)(20.756,0.000){0}{\usebox{\plotpoint}}
\put(365.29,537.00){\usebox{\plotpoint}}
\multiput(368,537)(20.756,0.000){0}{\usebox{\plotpoint}}
\put(386.04,537.00){\usebox{\plotpoint}}
\multiput(387,537)(20.756,0.000){0}{\usebox{\plotpoint}}
\multiput(396,537)(20.756,0.000){0}{\usebox{\plotpoint}}
\put(406.80,537.00){\usebox{\plotpoint}}
\multiput(414,537)(20.756,0.000){0}{\usebox{\plotpoint}}
\put(427.53,536.61){\usebox{\plotpoint}}
\multiput(433,536)(20.756,0.000){0}{\usebox{\plotpoint}}
\put(448.26,536.00){\usebox{\plotpoint}}
\multiput(451,536)(20.756,0.000){0}{\usebox{\plotpoint}}
\put(469.01,536.00){\usebox{\plotpoint}}
\multiput(470,536)(20.629,-2.292){0}{\usebox{\plotpoint}}
\multiput(479,535)(20.756,0.000){0}{\usebox{\plotpoint}}
\put(489.71,535.00){\usebox{\plotpoint}}
\multiput(498,535)(20.629,-2.292){0}{\usebox{\plotpoint}}
\put(510.41,534.00){\usebox{\plotpoint}}
\multiput(516,534)(20.756,0.000){0}{\usebox{\plotpoint}}
\put(531.14,533.39){\usebox{\plotpoint}}
\multiput(535,533)(20.756,0.000){0}{\usebox{\plotpoint}}
\put(551.82,532.13){\usebox{\plotpoint}}
\multiput(553,532)(20.756,0.000){0}{\usebox{\plotpoint}}
\multiput(562,532)(20.652,-2.065){0}{\usebox{\plotpoint}}
\put(572.52,530.94){\usebox{\plotpoint}}
\multiput(581,530)(20.756,0.000){0}{\usebox{\plotpoint}}
\put(593.20,529.64){\usebox{\plotpoint}}
\multiput(599,529)(20.652,-2.065){0}{\usebox{\plotpoint}}
\put(613.84,527.46){\usebox{\plotpoint}}
\multiput(618,527)(20.629,-2.292){0}{\usebox{\plotpoint}}
\put(634.47,525.17){\usebox{\plotpoint}}
\multiput(636,525)(20.652,-2.065){0}{\usebox{\plotpoint}}
\multiput(646,524)(20.629,-2.292){0}{\usebox{\plotpoint}}
\put(655.11,522.98){\usebox{\plotpoint}}
\multiput(664,521)(20.629,-2.292){0}{\usebox{\plotpoint}}
\put(675.54,519.49){\usebox{\plotpoint}}
\multiput(683,518)(20.629,-2.292){0}{\usebox{\plotpoint}}
\put(696.00,516.11){\usebox{\plotpoint}}
\multiput(701,515)(20.261,-4.503){0}{\usebox{\plotpoint}}
\put(716.38,512.36){\usebox{\plotpoint}}
\multiput(720,512)(20.261,-4.503){0}{\usebox{\plotpoint}}
\put(736.71,508.29){\usebox{\plotpoint}}
\multiput(738,508)(19.690,-6.563){0}{\usebox{\plotpoint}}
\put(756.75,503.05){\usebox{\plotpoint}}
\multiput(757,503)(20.261,-4.503){0}{\usebox{\plotpoint}}
\multiput(766,501)(19.690,-6.563){0}{\usebox{\plotpoint}}
\put(776.71,497.43){\usebox{\plotpoint}}
\multiput(784,495)(20.352,-4.070){0}{\usebox{\plotpoint}}
\put(796.72,492.09){\usebox{\plotpoint}}
\multiput(803,490)(19.690,-6.563){0}{\usebox{\plotpoint}}
\put(816.25,485.11){\usebox{\plotpoint}}
\multiput(821,483)(19.880,-5.964){0}{\usebox{\plotpoint}}
\put(835.68,477.92){\usebox{\plotpoint}}
\multiput(840,476)(19.690,-6.563){0}{\usebox{\plotpoint}}
\put(854.97,470.35){\usebox{\plotpoint}}
\multiput(858,469)(19.271,-7.708){0}{\usebox{\plotpoint}}
\put(874.10,462.29){\usebox{\plotpoint}}
\multiput(877,461)(18.144,-10.080){0}{\usebox{\plotpoint}}
\put(892.66,453.04){\usebox{\plotpoint}}
\multiput(895,452)(18.564,-9.282){0}{\usebox{\plotpoint}}
\put(911.13,443.60){\usebox{\plotpoint}}
\multiput(914,442)(18.144,-10.080){0}{\usebox{\plotpoint}}
\put(928.97,433.02){\usebox{\plotpoint}}
\multiput(932,431)(18.564,-9.282){0}{\usebox{\plotpoint}}
\put(946.94,422.71){\usebox{\plotpoint}}
\multiput(951,420)(17.270,-11.513){0}{\usebox{\plotpoint}}
\put(963.99,410.90){\usebox{\plotpoint}}
\multiput(969,407)(17.798,-10.679){0}{\usebox{\plotpoint}}
\put(981.17,399.31){\usebox{\plotpoint}}
\multiput(988,394)(16.383,-12.743){0}{\usebox{\plotpoint}}
\put(997.55,386.57){\usebox{\plotpoint}}
\put(1013.85,373.72){\usebox{\plotpoint}}
\multiput(1016,372)(15.513,-13.789){0}{\usebox{\plotpoint}}
\put(1029.45,360.04){\usebox{\plotpoint}}
\multiput(1034,356)(15.513,-13.789){0}{\usebox{\plotpoint}}
\put(1044.96,346.24){\usebox{\plotpoint}}
\put(1060.02,331.98){\usebox{\plotpoint}}
\multiput(1062,330)(14.676,-14.676){0}{\usebox{\plotpoint}}
\put(1074.70,317.30){\usebox{\plotpoint}}
\put(1089.38,302.62){\usebox{\plotpoint}}
\multiput(1090,302)(13.885,-15.427){0}{\usebox{\plotpoint}}
\put(1103.07,287.03){\usebox{\plotpoint}}
\put(1116.21,270.97){\usebox{\plotpoint}}
\multiput(1117,270)(13.962,-15.358){0}{\usebox{\plotpoint}}
\put(1129.78,255.29){\usebox{\plotpoint}}
\put(1136,247){\usebox{\plotpoint}}
\put(220,614){\usebox{\plotpoint}}
\put(220.00,614.00){\usebox{\plotpoint}}
\multiput(229,614)(20.756,0.000){0}{\usebox{\plotpoint}}
\put(240.76,614.00){\usebox{\plotpoint}}
\multiput(248,614)(20.756,0.000){0}{\usebox{\plotpoint}}
\put(261.51,614.00){\usebox{\plotpoint}}
\multiput(266,614)(20.756,0.000){0}{\usebox{\plotpoint}}
\put(282.27,614.00){\usebox{\plotpoint}}
\multiput(285,614)(20.756,0.000){0}{\usebox{\plotpoint}}
\multiput(294,614)(20.756,0.000){0}{\usebox{\plotpoint}}
\put(303.02,614.00){\usebox{\plotpoint}}
\multiput(313,614)(20.756,0.000){0}{\usebox{\plotpoint}}
\put(323.78,614.00){\usebox{\plotpoint}}
\multiput(331,614)(20.756,0.000){0}{\usebox{\plotpoint}}
\put(344.53,614.00){\usebox{\plotpoint}}
\multiput(350,614)(20.756,0.000){0}{\usebox{\plotpoint}}
\put(365.29,614.00){\usebox{\plotpoint}}
\multiput(368,614)(20.756,0.000){0}{\usebox{\plotpoint}}
\put(386.04,614.00){\usebox{\plotpoint}}
\multiput(387,614)(20.756,0.000){0}{\usebox{\plotpoint}}
\multiput(396,614)(20.629,2.292){0}{\usebox{\plotpoint}}
\put(406.74,615.00){\usebox{\plotpoint}}
\multiput(414,615)(20.756,0.000){0}{\usebox{\plotpoint}}
\put(427.50,615.00){\usebox{\plotpoint}}
\multiput(433,615)(20.629,2.292){0}{\usebox{\plotpoint}}
\put(448.20,616.00){\usebox{\plotpoint}}
\multiput(451,616)(20.652,2.065){0}{\usebox{\plotpoint}}
\put(468.91,617.00){\usebox{\plotpoint}}
\multiput(470,617)(20.629,2.292){0}{\usebox{\plotpoint}}
\multiput(479,618)(20.756,0.000){0}{\usebox{\plotpoint}}
\put(489.60,618.16){\usebox{\plotpoint}}
\multiput(498,619)(20.629,2.292){0}{\usebox{\plotpoint}}
\put(510.24,620.36){\usebox{\plotpoint}}
\multiput(516,621)(20.756,0.000){0}{\usebox{\plotpoint}}
\put(530.93,621.59){\usebox{\plotpoint}}
\multiput(535,622)(20.261,4.503){0}{\usebox{\plotpoint}}
\put(551.40,624.82){\usebox{\plotpoint}}
\multiput(553,625)(20.629,2.292){0}{\usebox{\plotpoint}}
\multiput(562,626)(20.652,2.065){0}{\usebox{\plotpoint}}
\put(572.04,627.01){\usebox{\plotpoint}}
\multiput(581,629)(20.261,4.503){0}{\usebox{\plotpoint}}
\put(592.34,631.26){\usebox{\plotpoint}}
\multiput(599,632)(20.352,4.070){0}{\usebox{\plotpoint}}
\put(612.76,634.84){\usebox{\plotpoint}}
\multiput(618,636)(20.261,4.503){0}{\usebox{\plotpoint}}
\put(632.85,639.95){\usebox{\plotpoint}}
\multiput(636,641)(20.352,4.070){0}{\usebox{\plotpoint}}
\put(652.87,645.29){\usebox{\plotpoint}}
\multiput(655,646)(19.690,6.563){0}{\usebox{\plotpoint}}
\put(672.56,651.85){\usebox{\plotpoint}}
\multiput(673,652)(19.880,5.964){0}{\usebox{\plotpoint}}
\multiput(683,655)(19.690,6.563){0}{\usebox{\plotpoint}}
\put(692.33,658.15){\usebox{\plotpoint}}
\multiput(701,662)(18.967,8.430){0}{\usebox{\plotpoint}}
\put(711.32,666.53){\usebox{\plotpoint}}
\multiput(720,670)(18.967,8.430){0}{\usebox{\plotpoint}}
\put(730.36,674.76){\usebox{\plotpoint}}
\multiput(738,679)(18.967,8.430){0}{\usebox{\plotpoint}}
\put(748.94,683.97){\usebox{\plotpoint}}
\multiput(757,688)(17.270,11.513){0}{\usebox{\plotpoint}}
\put(766.81,694.45){\usebox{\plotpoint}}
\multiput(775,699)(17.270,11.513){0}{\usebox{\plotpoint}}
\put(784.49,705.29){\usebox{\plotpoint}}
\put(801.63,716.93){\usebox{\plotpoint}}
\multiput(803,718)(17.270,11.513){0}{\usebox{\plotpoint}}
\put(818.13,729.45){\usebox{\plotpoint}}
\multiput(821,732)(17.004,11.902){0}{\usebox{\plotpoint}}
\put(834.52,742.13){\usebox{\plotpoint}}
\multiput(840,747)(15.513,13.789){0}{\usebox{\plotpoint}}
\put(850.03,755.92){\usebox{\plotpoint}}
\put(865.50,769.75){\usebox{\plotpoint}}
\multiput(868,772)(13.885,15.427){0}{\usebox{\plotpoint}}
\put(879.79,784.79){\usebox{\plotpoint}}
\put(894.01,799.90){\usebox{\plotpoint}}
\multiput(895,801)(13.962,15.358){0}{\usebox{\plotpoint}}
\put(907.79,815.41){\usebox{\plotpoint}}
\put(920.93,831.48){\usebox{\plotpoint}}
\multiput(923,834)(12.453,16.604){0}{\usebox{\plotpoint}}
\put(933.60,847.92){\usebox{\plotpoint}}
\put(946.34,864.27){\usebox{\plotpoint}}
\put(958.16,881.34){\usebox{\plotpoint}}
\multiput(960,884)(11.224,17.459){0}{\usebox{\plotpoint}}
\put(969.51,898.71){\usebox{\plotpoint}}
\put(981.28,915.79){\usebox{\plotpoint}}
\put(991.77,933.70){\usebox{\plotpoint}}
\put(1001.94,951.79){\usebox{\plotpoint}}
\put(1012.62,969.59){\usebox{\plotpoint}}
\put(1022.72,987.70){\usebox{\plotpoint}}
\put(1032.11,1006.21){\usebox{\plotpoint}}
\put(1041.39,1024.78){\usebox{\plotpoint}}
\put(1050.99,1043.18){\usebox{\plotpoint}}
\multiput(1053,1047)(7.708,19.271){0}{\usebox{\plotpoint}}
\put(1057,1057){\usebox{\plotpoint}}
\end{picture}

\end{center}
\caption{ 
Speed of light squared using the D234c action with coupling 
$C_3=1.2$ for pseudoscalar (P) and
vector (V) mesons, at $P/V=0.76$ (upper figure) and $P/V=0.6$
(lower figure).
}
\label{fig:csq12}
\end{figure}
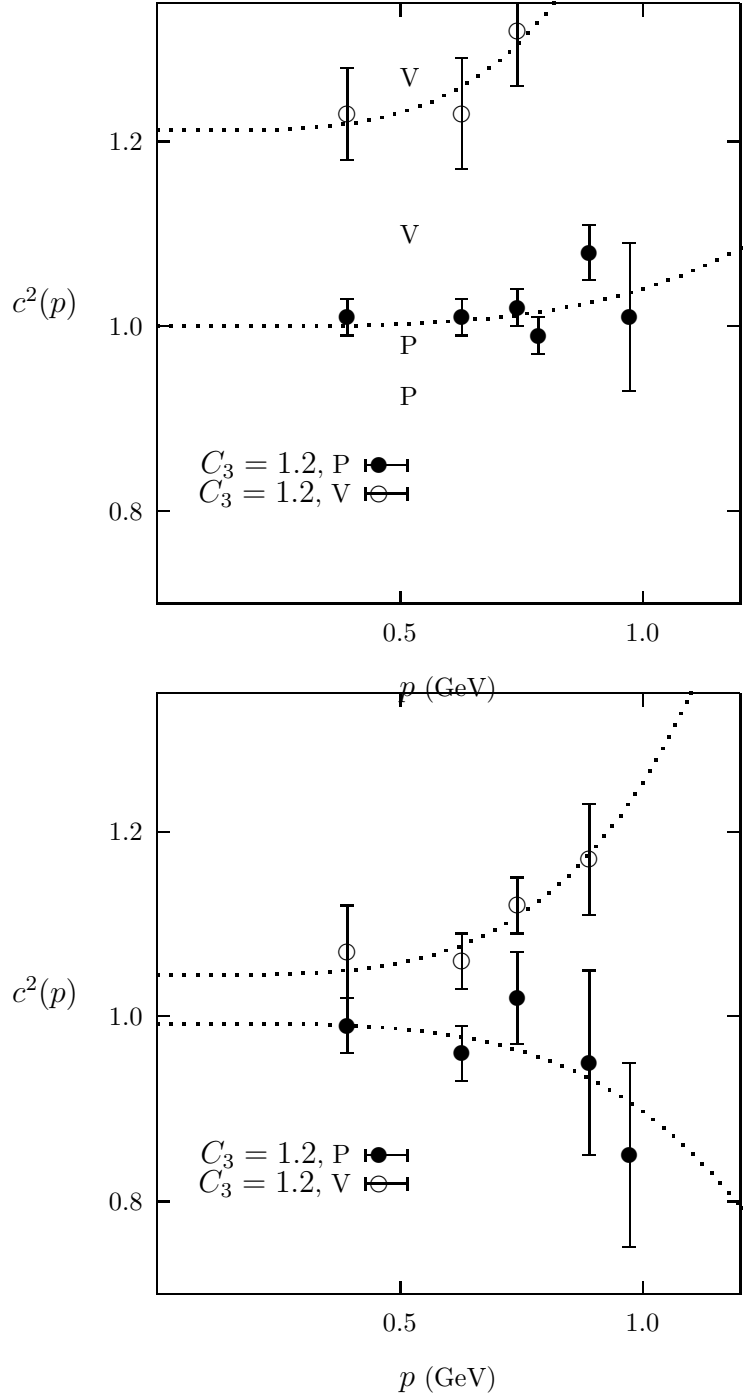

The $C_3$ term is the only $a^2$ correction that breaks rotational
invariance. Thus we can use $c^2$ to tune $C_3$~nonperturbatively. In our
0.4\,fm~simulation, we tuned $C_3$ to make the dispersion relation for
the lightest meson, the pseudoscalar, perfect at low momentum: when
\bea\label{eq:c3}
C_3 &=& 1.2 \\
&\approx& 1 + 0.5\alpha_s
\eea
we obtain the results shown in 
the top part of Figure~\ref{fig:csq12}. The
$c^2$ for the pseudoscalar is now
within $\pm2\%$ of $c^2\seq 1$ at least out to momenta of order $1.5/a$. 

It is noticeable in figure~\ref{fig:csq12} that tuning the
pseudoscalar $c^2$ has worsened the dispersion relation for the vector
meson ($c^2_V(p\!=\!0) = 1.21(5)$ from a quartic fit).  However, the
vector is 25\% heavier than the pseudoscalar and so should have
substantially larger $(a\,m_V)^n$~errors.  This is confirmed by a
rerun at lower quark mass, $P/V=0.6$, where the vector is lighter
(bottom part of Figure~\ref{fig:csq12}).
There we see that chosing $C_3=1.2$ works for both pseudoscalar and
vector: the vector now has $c^2(p\!=\!0) = 1.04(4)$.
In table \ref{tab:c2Csw} we see that, to within statistical errors,
the pseudoscalar meson
dispersion relation is insensitive 
to $C_F$, confirming that $C_3$ can be tuned independently of $C_F$.
(We ignore the vector meson because, as noted above, at $a=0.4~\fm$
and $P/V=0.7$ it is too heavy for its dispersion relation to be a
reliable indicator of the Lorentz-violating errors.)

From these results it is clear that for the D234c action, as for SW, 
the hadron dispersion relation is more sensitive to
finite-$a$ errors than the hadron mass. 
At $P/V=0.7$
on an isotropic $0.4~\fm$ lattice we get a perfectly sensible
vector mass (Fig.~\ref{fig:phi70}), but the dispersion relation
is beginning to break down due to $\O((am)^2)$ errors.
An anisotropic lattice would extend D234c's range to much higher
quark masses at the same spatial lattice spacing, because the
errors due to the quark mass are of order $(a_t m)^2$.

\subsection{Redundant terms}

As a final test of the D234c action, we varied the parameter~$r$. The
$r$-dependent terms in the D234c action are, by design, approximately
redundant, and therefore varying~$r$ should have little effect on the
spectrum. Furthermore the extent to which this is true gives us an
indication of the size of the finite-$a$ errors associated with the
$r$~terms, including the dominant correction, the clover term. Our
results are summarized in Table~\ref{tab:r-test}. As expected, the
variation with $r$ is much smaller with D234c than with the SW or Wilson
actions\,---\,the $r$~terms are more highly corrected in the former.
This ``$r$-test'' suggests that $r$-dependent finite-$a$ errors in the
static masses are of
order 3--4\% at 0.25\,fm, and 6\% at 0.4\,fm for D234c. Our discussion
above shows that these errors could be almost entirely due to
radiative corrections to the clover coupling~$C_F$. 

\def\phXXX{\protect\phantom{XXX}}
\def\st{\rule[-1.5ex]{0em}{4ex}} 
\begin{table} 
\begin{tabular}{lllll}
\hline
\st Action     & Wilson & SW  & \multicolumn{2}{l}{\phXXX D234c}\\
\st $a$ (fm) &  0.4   & 0.4 & 0.25     &  0.4           \\
\hline
\st P & 16.0(7)\%  & 8(2)\% & 1.0(1)\% & 2.0(7)\% \\
\st V & 15.2(11)\% & 8(2)\% & 1.3(2)\% & 2.0(5)\% \\
\st N & 9.0(14)\%  & 5(2)\% & 0.7(2)\% & 1.1(6)\% \\
\st D & 9.0(20)\%  & 8(2)\% & 0.9(2)\% & 1.5(6)\%  \\
\hline
\end{tabular}
\caption{
Percentage change in hadron masses 
when the redundant coefficient $r$ is changed from
1 to 2/3.  Scaling violation due to redundant terms at $r=1$ is therefore 
expected to be roughly 3 times this.
}
\label{tab:r-test}
\end{table}

\def\order{{\cal O}}
\section{Conclusions}
Our studies show that at tree level the mean-link tadpole-improved
D234c action is accurate to about 10\% in hadronic masses, even for
lattice spacings as large as~0.4\,fm and meson masses as large as
$2/a\!\approx\!1$\,GeV. The meson dispersion relation ($c^2(\pv)$) is
similarly accurate out to three-momenta of order
$1.5/a\!\approx\!750$\,MeV. It is markedly superior to the SW action
in the limit of large masses or momenta.

Our analysis indicates that the leading source of error in the hadron
masses is the radiative correction, beyond tadpole-improved
tree-level, to the clover coefficient~$C_F$, which enters in the
$\order(a)$~correction. Radiative corrections to other terms
contribute probably less than~5\% at $a=0.4~\fm$, while uncorrected
$a^4$~errors are probably no more than a few percent.
Thus the errors in the hadron spectrum in our
0.4\,fm~simulation can almost certainly be reduced to less than 10\% 
by correcting~$C_F$.

We estimated the size of the quantum corrections to $C_F$ by tuning
$C_F$ until our hadron masses, computed with $a=0.25$\,fm and~0.4\,fm,
agreed with a quadratic $a\seq0$ extrapolation of SCRI's 
(tadpole-improved) SW data. We
found that the quantum corrections to~$C_F$ are of the right size to
be perturbative.
The perturbative prediction for $C_F$ is not yet
known, but we may compare with the clover coefficient in the SW action
which, for Wilson's (unimproved) gluon action, has the perturbative
expansion $1+0.46\,\alpha_s$ after mean-link
tadpole-improvement\,\cite{gplTsukuba},
which agrees with non-perturbative estimates to within 
a reasonably sized $\O(\al^2)$ error.

We also demonstrated how to nonperturbatively tune the leading
$\order(a^2)$~correction, the $C_3$~term in the D234c action, by using
restoration of rotational invariance. The correction to tadpole-improved
tree-level is small, and is unimportant at the few percent level
for~$a\lesssim 0.4$\,fm.  

As expected, all errors were much smaller at $a\seq0.25$\,fm, and
radiative corrections were much less important. With a properly tuned
clover coefficient, $C_F$, D234c should 
give (static and kinetic) hadron masses accurate to within a 
couple of percent at this relatively large lattice spacing. 

In the near future we will complete perturbative calculations of $C_F$
and $C_3$ to compare with our nonperturbative tunings; nonperturbative
tuning of $C_F$ using PCAC~restoration is another possibility
\cite{TimRob}.  We are also examining currents, form factors and other
quantities with the D234c action. Perhaps most importantly, we are
continuing our previous work on D234c for anisotropic lattices. This
simulation technology is applicable to a range of phenomenological
applications (hadronic spectra and decay constants, relativistic
heavy-quark physics, high-momentum form factors), and we anticipate
being able to obtain useful results with spatial lattice spacings in the range 
0.2--0.4\,fm.


\appendix
\section{Galilean Quarkonium}
\label{app:lattspacing}

As explained in Section \ref{sec:lattspac}, Galilean quarkonium 
is a convenient fictitious state for the evaluation of ratios
of lattice spacings.
It is simulated using the conventional NRQCD quark action, without
relativistic corrections:
\beq
G({\vec x},t+a) = 
\Bigl(1 - a{H_0\over 2n}\Bigr)^n
\Bigl(1 - a{\de H\over 2}\Bigr)
U^\ad_{{\vec x},t}
\Bigl(1 - a{\de H\over 2}\Bigr)
\Bigl(1 - a{H_0\over 2n}\Bigr)^n   G({\vec x},t)
\eeq
where
\beq
H_0 = - { {\bf \De}^{(2)} \over 2M}
\eeq
\beq
\de H = - {a\over 4n} { ({\bf \De}^{(2)})^2 \over 4M^2}
 + {a^2\over 24M} \De^{(4)}
\eeq
For details of the lattice derivatives and fields, see \cite{NRQCD}.

For this paper, we needed the ratios of our two lattice spacings
to each other, and to SCRI's $\be=7.4$ improved glue.
We therefore performed Galilean charmonium simulations using all 3
types of glue, at a range of quark masses. The quark mass is
characterized by the dimensionless quantity
${M_{\rm kin}/ (P-S)}$. $M_{\rm kin}$ is the kinetic mass
of the quarkonium $S$-state, obtained from 
measurements of the energy of the $S$-state 
at rest and with the smallest momentum on the lattice.
This quantity is 6.8 for charmonium, so our Galilean quarkonium
states were mostly lighter than charmonium.

Our raw measurements of the $P-S$ splitting are given in table
\ref{tab:galquarkonium}, and the resultant lattice spacing ratios are
in table \ref{tab:galileo}.  We see that there is no dependence of
the ratios on the quark mass within statistical errors, so that
finite-$a$ errors in our ratios are negligible.

Using a potential model for
Galilean quarkonium, we find that the radius of the meson changes
by $25-30\%$ over this range of quark masses, so any uncorrected $a^4$
errors would double in size.
Interestingly, the potential model gives similar sensitivities
of $P-S$ splitting to quark mass to those measured on the
lattice (last line of table \ref{tab:galquarkonium}).

\clearpage

%
%
\def\st{\rule[-1.5ex]{0em}{4ex}} 
\begin{table}[htb]
\begin{tabular}{lllll}
\hline
\rule[-2.5ex]{0em}{6.4ex} 
 &\multicolumn{4}{c}{$\dsp{M_{\rm kin}\over (P-S)}$} \\
\cline{2-5}
\st $\beta$ & 3.6 & 4.0 & 5.2 & 7.2 \\
\hline
\st 1.157      &  1.0846(33) & 1.0545(60) & 1.0124(70) & 0.9667(80) \\
\st 1.719      &  0.6863(35) & 0.6729(50) & 0.6317(80) & 0.5990(51) \\
\st 7.4 (SCRI) &  0.6638(36) & 0.6513(67) \\

\rule[-2.5ex]{0em}{6ex} $\dsp {d\ln(P-S)\over d M_{\rm kin}/(P-S)}$
               & -0.06(1)    & -0.05(1)   & -0.04(2)   & -0.02(1) \\
\hline
\end{tabular}
\caption{ Measurements of $a(P-S)$ for Galilean quarkonium
at various kinetic quarkonium masses $M_{\rm kin}$. Also (final line),
sensitivity of $P-S$ to quark mass.
}
\label{tab:galquarkonium}
\end{table}

\def\st{\rule[-1.5ex]{0em}{4ex}} 
\begin{table}[htb]
\begin{tabular}{cllll}
\hline
\rule[-2.5ex]{0em}{6.4ex} 
 &\multicolumn{4}{c}{$\dsp{M_{\rm kin}\over (P-S)}$} \\
\cline{2-5}
\st  & 3.6 & 4.0 & 5.2 & 7.2 \\
\hline
\rule[-2.8ex]{0em}{7ex} $\dsp {a(\be\!=\!1.157) \over a(\be\!=\!1.719)}$
 & 0.633(4) & 0.638(6) & 0.624(9) & 0.620(8) \\
\rule[-2.8ex]{0em}{7ex} 
 $\dsp {a(\be\!=\!1.157) \over a({\rm SCRI},\be\!=\!7.4)}$
 & 0.612(4) & 0.618(7) \\
\rule[-2.8ex]{0em}{7ex} 
 $\dsp {a(\be\!=\!1.719) \over a({\rm SCRI},\be\!=\!7.4)}$
 & 0.967(7) & 0.968(12) \\
\hline
\end{tabular}
\caption{ Ratios of lattice spacings calculated using Galilean quarkonium,
for a variety of quarkonium masses. Our glue ($\be= 1.157, 1.719$) is
mean-link tadpole-improved, SCRI's ($\be= 7.4$) is plaquette tadpole-improved.
}
\label{tab:galileo}
\end{table}

\clearpage

\section{Data Tables}
\label{app:tables}


\def\st{\rule[-1.5ex]{0em}{4ex}} 
\begin{table}[htb]
\begin{tabular}{llllll}
\hline
\st $\beta$ & lattice & $am_{\rm quark}$ & $a(P-S)$ & $a$ (fm) & $a^{-1}$ (MeV) \\
\hline
\st 1.157 & $5^3 \times 10$ & 2.75 & 0.925(7)  & 0.400(4) & 495(4) \\
\st 1.719 & $8^4$           & 1.80 & 0.578(10) & 0.249(5) & 790(10) \\
\hline
\end{tabular}
\caption{
Determination of lattice spacing from NRQCD calculation of
spin-averaged charmonium $P-S$ splitting, $P-S = 458~\MeV$.
The lattice size and bare quark mass are given.
}
\label{tab:charmglue}
\end{table}


\def\phXXX{\protect\phantom{XXX}}
\def\st{\rule[-1.5ex]{0em}{4ex}} 
\begin{table}[htb]
\begin{tabular}{llll@{\phXXX}lll}
\hline
\st & \multicolumn{3}{l}{D234c, $am_{\rm quark}=0.640$ } & 
      \multicolumn{3}{l}{SW, $am_{\rm quark}=0.594$ } \\
\hline
\st & $6^3$ & $7^3$ & $8^3$ & $6^3$ & $7^3$ & $8^3$ \\
\hline
\st P & 0.901(5)  & 0.917(6)  & 0.908(4) & 0.939(6)  & 0.955(6)  & 0.945(4) \\
\st V & 1.293(9)  & 1.312(12) & 1.304(8) & 1.332(8)  & 1.350(15) & 1.333(8) \\
\st N & 1.950(20) & 1.955(15) & 1.950(25) & 1.960(40) & 1.990(20) & 1.958(30)\\
\st D & 2.210(20) & 2.210(20) & 2.220(20) & 2.250(20) & 2.220(40) & 2.230(25) \\
\hline
\end{tabular}
\caption{
Finite volume errors:
Hadron masses in lattice units at $P/V\approx 0.7$, $a=0.25~\fm$, measured on
$6^3\times 20$, $7^3\times 20$, and $8^3\times 20$ lattices.
}
\label{tab:vol}
%
\end{table}


\def\phXXX{\protect\phantom{XXX}}
\def\st{\rule[-1.5ex]{0em}{4ex}} 
\begin{table}[htb]
\begin{tabular}{l@{\phXXX}ll@{\phXXX}ll}
\hline
\st & \multicolumn{2}{l}{$a=0.25~\fm$} & \multicolumn{2}{l}{$a=0.40~\fm$} \\
\hline
\st & D234c & SW & D234c & SW \\
\hline
\st quark & 0.68      & 0.63      & 1.002     & 1.012 \\
\st P/V   & 0.756(5)  & 0.762(6)  & 0.756(4)  & 0.763(3) \\
\st P     & 1.051(4)  & 1.085(4)  & 1.583(3)  & 1.497(7) \\
\st V     & 1.391(7)  & 1.424(7)  & 2.095(10) & 1.960(10)\\
\st N     & 2.100(30) & 2.150(24) & 3.220(15) & 3.000(20) \\
\st D     & 2.340(30) & 2.370(20) & 3.540(30) & 3.260(20) \\
\hline
\end{tabular}
\caption{
Hadron masses in lattice units at $P/V \approx 0.76$.
For $a=0.40~\fm$ we used a $5^3\times 18$ 
lattice, for $a=0.25~\fm$ an $8^3\times 20$.
Physical spatial volume is the same.
}
\label{tab:hadrons76}
\end{table}

\def\phXXX{\protect\phantom{XXX}}
\def\st{\rule[-1.5ex]{0em}{4ex}} 
\begin{table}[htb]
\begin{tabular}{l@{\phXXX}ll@{\phXXX}ll}
\hline
\st & \multicolumn{2}{l}{$a=0.25~\fm$} & \multicolumn{2}{l}{$a=0.40~\fm$} \\
\hline
\st & D234c & SW & D234c & SW \\
\hline
\st quark & 0.640     & 0.594     & 0.943     & 0.970 \\
\st P/V   & 0.694(3)  & 0.708(8)  & 0.703(4)  & 0.699(8) \\
\st P     & 0.902(3)  & 0.945(4)  & 1.371(3)  & 1.313(15) \\
\st V     & 1.300(4)  & 1.333(8)  & 1.950(10) & 1.878(10)\\
\st N     & 1.925(25) & 1.958(30) & 2.960(20) & 2.850(20) \\
\st D     & 2.170(20) & 2.230(25) & 3.380(20) & 3.150(30) \\
\hline
\end{tabular}
\caption{
Hadron masses in lattice units at $P/V \approx 0.70$.
For $a=0.40~\fm$ we used a $5^3\times 18$ lattice, for $a=0.25~\fm$ 
an $8^3\times 20$. Physical spatial volume is the same.
}
\label{tab:hadrons70}
\end{table}


\def\phXXX{\protect\phantom{XXX}}
\def\st{\rule[-1.5ex]{0em}{4ex}} 
\begin{table}[htb]
\begin{tabular}{l@{\phXXX}ll@{\phXXX}ll}
\hline
\st & \multicolumn{2}{l}{ $c^2(P)$} & \multicolumn{2}{l}{$c^2(V)$} \\
\hline
\st mom &  $\Csw=1$ & $\Csw=0.72$ &  $\Csw=1$ & $\Csw=0.72$ \\
\hline
\st 001 & 0.95(2) &  0.99(2)  & 1.05(5) & 0.93(5) \\
\st 100 & 0.94(4) &  0.91(3)  & 1.08(3) & 0.91(3) \\
\st 101 & 0.94(4) &  0.96(2)  & 1.16(4) & 0.96(4) \\
\st 002 & 0.90(2) &  0.93(4)  &  & \\
\st 110 & 0.90(3) &  0.89(5)  & 1.30(10) & 0.92(6)\\
\st 111 & 0.84(5) &  0.98(9) &         & \\
\hline
\end{tabular}
\caption{
Sensitivity of $c^2$ to $\Csw$:
Hadron $c^2$ at $P/V=0.76$, $5^2\times 8 \times 18$, $a=0.4~\fm$ lattice.
The vector particle's $c^2$ shows definite sensitivity to
a change in $\Csw$ ($P/V$ kept constant)
}
\label{tab:c2Csw}
\end{table}

\def\phXXX{\protect\phantom{XXX}}
\def\st{\rule[-1.5ex]{0em}{4ex}} 
\begin{table}[htb]
\begin{tabular}{l@{\phXXX}lll@{\phXXX}lll}
\hline
\st & \multicolumn{3}{l}{ $c^2(P)$} & \multicolumn{3}{l}{$c^2(V)$} \\
\hline
\st mom & SW & $C_3=1$ & $C_3=1.2$  & SW     & $C_3=1$ & $C_3=1.2$ \\
\hline
\st 001 & 0.632(24) & 0.96(2) &  1.01(2) & 0.421(16) & 1.05(5) & 1.23(5) \\
\st 100 & 0.550(13) & 0.93(3) &  1.01(2) & 0.342(13) & 1.08(3) & 1.23(6) \\
\st 101 & 0.560(14) & 0.93(3) &  1.02(2) & 0.355(14) & 1.17(4) & 1.32(6) \\
\st 002 & 0.510(16) & 0.90(2) &  0.99(2) &           &         & \\
\st 110 & 0.516(20) & 0.90(3) &  1.08(3) & 0.345(13) & 1.28(9) & \\
\st 111 & 0.505(60) & 0.84(5) &  1.01(8) &           &         & \\
\hline
\end{tabular}
\caption{
Sensitivity of $c^2$ to $C_3$, the coeff of $\De^{(3)}$. 
Tree-level value is $C_3 =1$, SW has $C_3=0$.
The table shows hadron $c^2$ at $P/V=0.76$ on 
a $5^2\times 8\times 18$, $a=0.4~\fm$ lattice.
From the pseudoscalar $c^2$, $C_3=1.2$ appears to
be the results that would follow from non-perturbative tuning.
}
\label{tab:c2_b}
\end{table}

\def\phXXX{\protect\phantom{XXX}}
\def\st{\rule[-1.5ex]{0em}{4ex}} 
\begin{table}[htb]
\begin{tabular}{l@{\phXXX}ll@{\phXXX}ll}
\hline
\st & \multicolumn{2}{l}{ $c^2(P)$} & \multicolumn{2}{l}{$c^2(V)$} \\
\hline
\st mom & SW & D234c & SW & D234c \\
\hline
\st 100 & 0.79(3) & 0.99(3) & 0.68(6) & 0.97(5) \\
\st 110 & 0.76(4) & 0.95(4) & 0.65(3) & 0.96(4) \\
\st 111 & 0.84(9) & 0.98(9) & \\
\st 200 & 0.70(9) & 0.94(8) & \\
\hline
\end{tabular}
\caption{
Hadron $c^2$ at $P/V=0.76$, on an $8^3 \times 20$, $a=0.25~\fm$ lattice.
}
\label{tab:c2_a0.25}
\end{table}

\end{document}